\documentclass[preprint2]{proto}
\usepackage{times}
\usepackage{xcolor}
\usepackage{multirow}
\usepackage{booktabs} \newcommand{\ra}[1]{\renewcommand{\arraystretch}{#1}}\ra{1.3}
\usepackage{tabularx} 
\usepackage{tabulary}
\usepackage{colortbl}
\usepackage{calc}
\usepackage{rotating}
\usepackage{graphicx}
\usepackage{subfig}

\newcommand{\cf}{cf.\ }

\newcommand{\eg}{e.g., }

\newcommand{\ie}{i.e., }

\newcommand{\sect}[1]{Section \ref{s:#1}}
\newcommand{\secttwo}[2]{Sections \ref{s:#1}--\ref{s:#2}}

\newcommand{\eqn}[1]{Eq.\ (\ref{e:#1})}

\newcommand{\fig}[1]{Fig.\ \ref{fig:#1}}

\newcommand{\Fig}[1]{Figure \ref{fig:#1}}

\newcommand{\tbl}[1]{Table \ref{t:#1}}

\newcommand{\code}[1]{\texttt{#1}}

\newcommand{\bvec}[1]{\mathbf{#1}}           
\newcommand{\bsym}[1]{\mbox{\boldmath $#1$}} 

\newcommand{\hide}[1]{} 

\let\OLDthebibliography\thebibliography
\renewcommand\thebibliography[1]{
  \OLDthebibliography{#1}
  \setlength{\parskip}{0pt}
  \setlength{\itemsep}{0pt plus 0.3ex}
}

\begin{document}

\title{\textbf{\LARGE Asteroid Surface Geophysics}}

\author {\textbf{\large Naomi Murdoch}}
\affil{\small\em Institut Sup{\'e}rieur de l'A{\'e}ronautique et de l'Espace (ISAE-SUPAERO)}

\author {\textbf{\large Paul S{\'a}nchez}}
\affil{\small\em University of Colorado Boulder}

\author {\textbf{\large Stephen R. Schwartz}}
\affil{\small\em University of Nice-Sophia Antipolis, Observatoire de la C{\^o}te d'Azur}

\author {\textbf{\large Hideaki Miyamoto}}
\affil{\small\em University of Tokyo}

\begin{abstract}
\begin{list}{ } {\rightmargin 0.5in}
\baselineskip = 11pt
\parindent=1pc
{\small 
{
The regolith-covered surfaces of asteroids preserve records of geophysical processes that have occurred both at their surfaces and sometimes also in their interiors. As a result of the unique micro-gravity environment that these bodies posses, a complex and varied geophysics has given birth to fascinating features that we are just now beginning to understand.  The processes that formed such features were first hypothesised through detailed spacecraft observations and have been further studied using theoretical, numerical and experimental methods that often combine several scientific disciplines.  These multiple approaches are now merging towards a further understanding of the geophysical states of the surfaces of asteroids.  
In this chapter we provide a concise summary of what the scientific community has learned so far about the surfaces of these small planetary bodies and the processes that have shaped them. We also discuss the state of the art in terms of experimental techniques and numerical simulations that are currently being used to investigate regolith processes occurring on small-body surfaces and that are contributing to the interpretation of observations and the design of future space missions.
\\~\\~}}

\end{list}
\end{abstract}  

\vskip 1cm

\section{\textbf{INTRODUCTION}}

Before the first spacecraft encounters with asteroids, many scientists assumed that the smallest asteroids were all monolithic rocks with a bare surface, although, there had been a few articles suggesting possible alternative surface properties and internal structures \citep[\eg][]{dollfus1977,housen1979,michel2001,harris2006}. Given the low gravitational acceleration on the surface of an asteroid, it was thought that regolith formation would not be possible; even if small fragments of rock were created during the impact process nothing would be retained on the surface \citep[\eg][]{chapman1976}. {{ However, the NASA Galileo, NEAR-Shoemaker (hereafter simply NEAR) and the JAXA Hayabusa space missions revealed a substantial regolith covering (951) Gaspra, (243) Ida, (433) Eros \citep{sullivan2002,robinson2002} and (25143) Itokawa \citep{fujiwara2006}. } } In addition to finding each of these bodies to be regolith-covered, there is strong evidence that this regolith has very complex and active dynamics. In fact, it was due to the NEAR observations of Eros that the local gravity was first understood to be of importance to asteroid surface processes \citep{robinson2002}. The importance of gravity for regolith dynamics was emphasised even further when the first images were received from the Hayabusa probe.

Over the course of these space missions and others a wide range of geological features have been observed on the surfaces of asteroids and other small bodies such as the nucleus of comet 103P/Hartley 2 \citep{thomas2013}. However, we do not have direct access to the properties of the granular material that led to these features.  Although constitutive equations exist for granular interactions on Earth, the inferred scaling to the gravitational and environmental conditions on other planetary bodies such as asteroids is currently untested.  Understanding the dynamics of granular materials in the small-body gravitational environment is vital for the interpretation of their surface geology and is also critical for the design and/or operation of any device planned to interact with their regolith-covered surfaces.  

{
Regolith was originally defined as ``a layer of fragmented debris of relatively low cohesion which overlies a more coherent substratum" \citep{shoemaker1968}, although, this definition runs into difficulties when there is no clear interface separating the fragmented debris and the coherent substrate \citep{robinson2002}.   Here we will use the term regolith to describe, in general terms, the ``loose unconsolidated material that comprises the upper portions of the asteroid" \citep[as defined in][]{robinson2002}.  However, we note that self-gravitating aggregates like Itokawa,  often referred to as ``rubble piles'' \citep{richardson2002},  are composed of rubble - boulders of the order of tens of metres and less - held together by gravity and cohesive forces instead of being a monolithic body \citep{fujiwara2006}.  As such they are essentially made of regolith throughout. Therefore, although not discussed in this chapter, understanding how granular materials behave in these extremely low-gravity environments can also improve our understanding of the interiors of these bodies. }

This chapter will start by presenting our current knowledge of the surfaces of asteroids (433) Eros, (25143) Itokawa, (21) Lutetia and (4) Vesta. After a short introduction to granular materials, we will then introduce the unique asteroid environment and suggest how this may influence the regolith dynamics.  Next, we discuss in detail the underlying physical mechanisms behind the geological processes observed to occur on the surfaces of asteroids. Finally, a discussion of the experimental techniques that can be used to simulate the asteroid environment and the recent advances in modelling regolith dynamics is provided. 

\section{\textbf{IN-SITU OBSERVATIONS OF ASTEROID SURFACES}}\label{s:Observations}

In this section we will briefly discuss the in-situ observations of four asteroids: (433) Eros, (25143) Itokawa, (21) Lutetia and (4) Vesta. For more detailed reviews about these bodies the readers are referred to \cite[{\it Yoshikawa et al.}; {\it Barucci et al.}; {\it Russell et al. (all this volume)} and][]{cheng2002b}. Additionally, detailed reviews of the geology of other asteroids such as (951) Gaspra, (243) Ida and (253) Mathilde are available elsewhere \citep[\eg][]{carr1994, sullivan1996,thomas1999}. 

  \subsection{Asteroid (433) Eros}\label{s:Eros}

\begin{figure}[htbp]
\begin{center}
\includegraphics[scale=0.35]{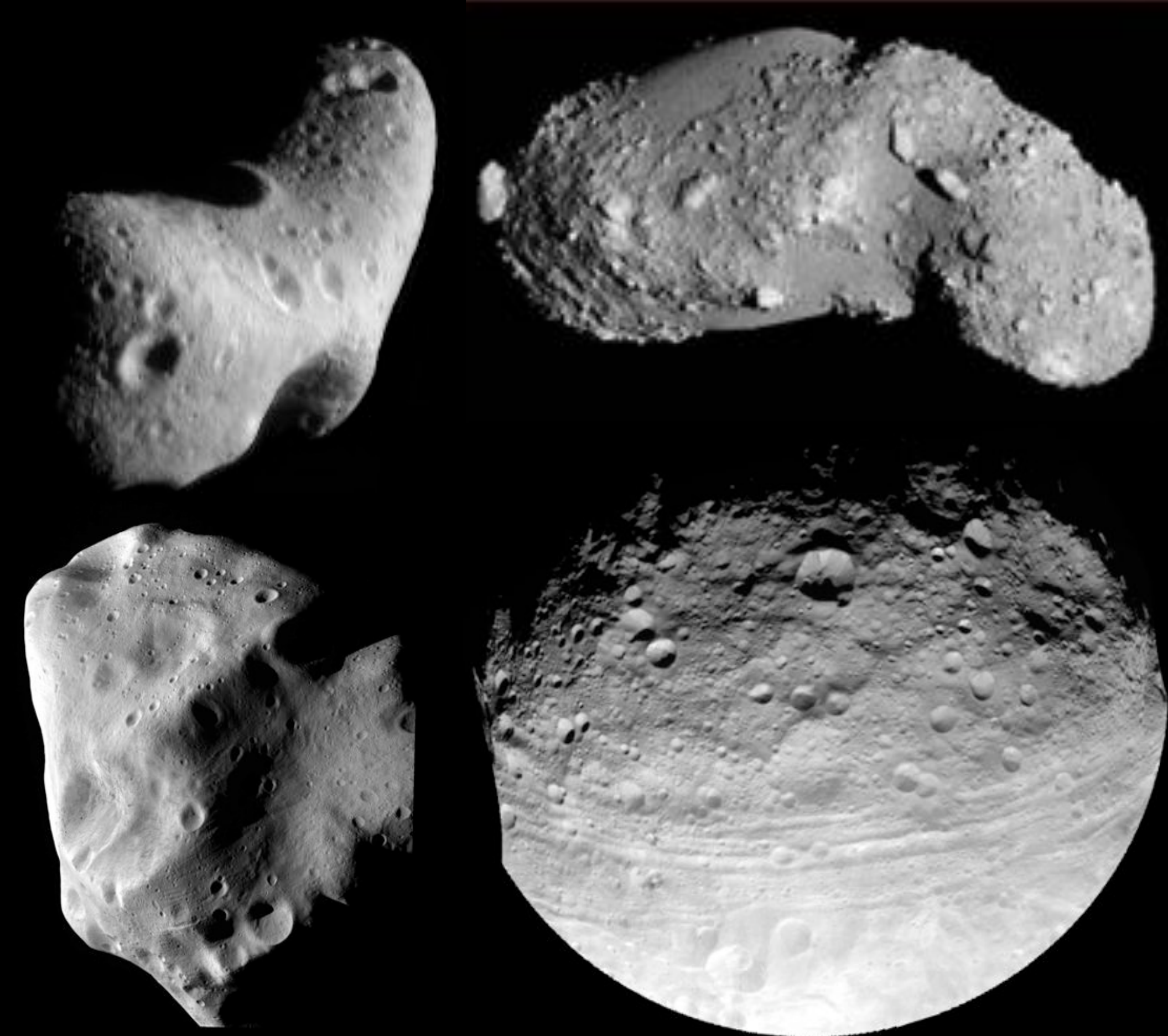}
\caption{Global images of asteroids (433) Eros (top left), (25143) Itokawa (top right), (21) Lutetia (bottom left) and (4) Vesta (bottom right). Image credits: NEAR/NASA, Hayabusa/JAXA, Rosetta/ESA, Dawn/NASA, respectively.}
\label{fig:Asteroids}
\end{center}
\end{figure}

(433) Eros (hereafter simply Eros; \fig{Asteroids},\tbl{asteroids}), the second largest near-Earth asteroid (NEA), shows a subdued, gently undulating and complex regolith-covered surface, characterised by abundant, but not uniformly distributed, ejecta blocks and conspicuously degraded craters \citep{veverka2000,veverka2001b,cheng1997}.  The effective topography on Eros has a range of about 2 km and the slopes, calculated relative to the local gravity vector, vary over the surface of the asteroid \citep[for an explanation of how elevation is defined on irregular bodies see Section 5 of][]{cheng2002c} with an average slope of $\sim8^\circ$ to 10$^\circ$ \citep{zuber2000, thomas2002}. 
 
\paragraph{Evidence of regolith motion on Eros - }

\begin{figure}[hbp]
\begin{center}
\includegraphics[scale=0.32]{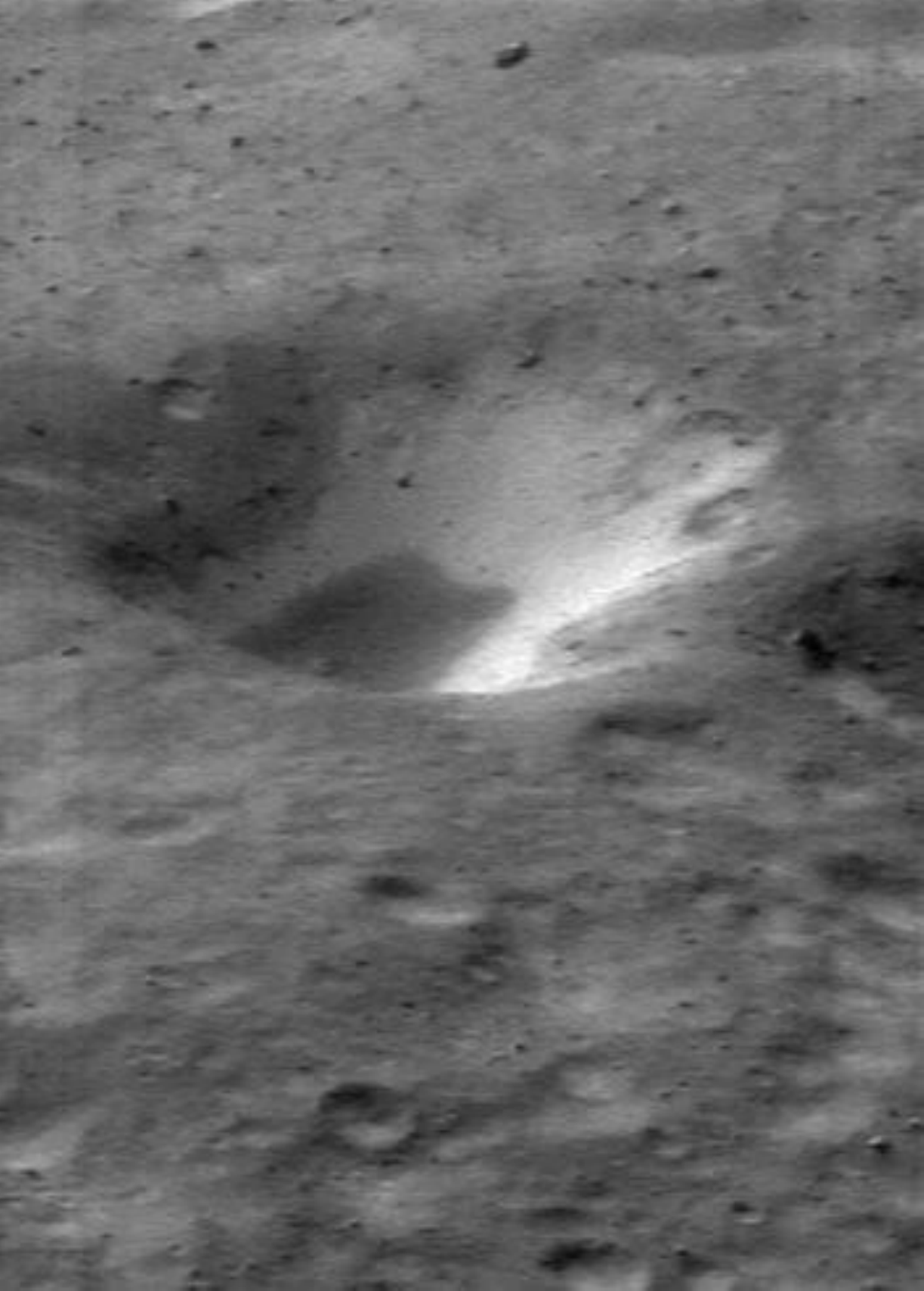} 
\includegraphics[scale=0.25]{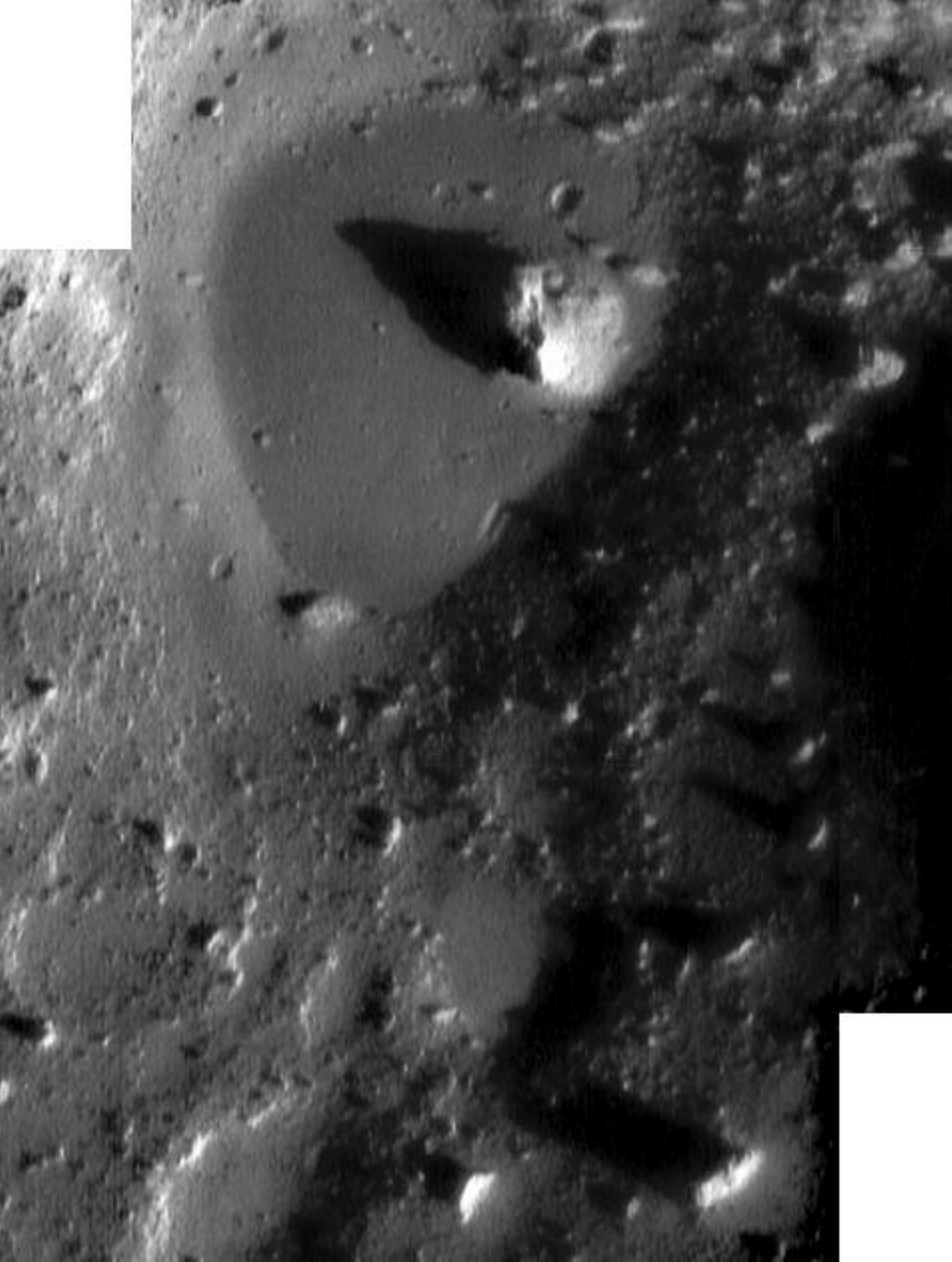}
\caption{Evidence of regolith transport on Eros - LEFT: Bright, freshly exposed material on a large crater wall, as the darker material moves downslope (PIA03134); RIGHT: An example of a dust pond (2001\_028\_5\_eros.png from {\it http://ser.sese.asu.edu/near.html}).}
\label{fig:ErosRegolithTransport}
\end{center}
\end{figure}

In general, Eros is very bland in terms of colour and albedo variations. However, visible variations, such as several bright features typically with sharp boundaries, can be seen in regions that have steep slopes \citep[see \fig{ErosRegolithTransport};][]{veverka2000,robinson2001,cheng2002b,mantz2004,murchie2002}.  As all the NEAR data indicate global compositional homogeneity, the brighter surfaces imply freshly exposed material that has not yet been subjected to space weathering \cite[{\it Brunetto et al. (this volume)} and][]{chapman2004}. In contrast, dark soils are typically located at the bases of bright streaks and display both diffuse and sharp boundaries \citep{thomas2002,riner2008}. These observations, and morphological data, indicate that the bright streaks are the results of preferential downslope movement or a landslide of mature regolith, revealing immature material beneath \citep[\eg][]{robinson2002,thomas2002,riner2008,murchie2002}.  

Indeed, on closer inspection, accumulations of granular material that have been gravitationally transported away from topographic highs can be seen on Eros \citep[\fig{ErosRegolithTransport}; ][]{thomas2002, veverka2001a, robinson2002}.  These granular deposits appear to result from low momentum downslope movements and some observations suggests that mobilised regolith may even be halted by frictional or other effects before reaching the foot of the slope \citep{mantz2004, thomas2002}.  Downslope motion has also been observed on slopes that are well below the expected angle of repose for granular materials.  Whether this indicates the necessity for a triggering mechanism or not is a subject currently under debate \citep{cheng2002b, holsapple2013}.

\paragraph{Craters and crater morphology on Eros - }

Further evidence for regolith motion is that, despite the large number of craters on the surface of Eros, there is a deficiency of small ($<$2 km diameter) craters \citep{veverka2000, veverka2001b}.   As there are sufficient projectiles in near-Earth space to produce small craters there must, therefore, be a process that either covers or erodes small craters on Eros \citep{veverka2001}.  It has been suggested that impact-induced seismic shaking (see \sect{Processes}), which causes the regolith to move, may erase small crater features and thus explain their paucity compared to predictions of dynamical models of projectile populations \citep[\eg][]{richardson2004, michel2009}.  However, alternative degradation mechanisms have also been suggested including micro-cratering and thermal creep \citep{cheng2002b}.  See Marchi et al. (this volume) for a detailed discussion of cratering on asteroids.

Additional evidence of resurfacing and modification is visible in the interiors and the subdued rims of several craters \citep{robinson2002,zuber2000}. The depth-to-diameter ratio of craters on Eros is, on average, $\sim$0.13, but the freshest and youngest craters approach lunar values of $\sim$0.2 \citep{robinson2002}.  Many of the topographic lows are filled with deposits of fine granular material \citep[\fig{ErosRegolithTransport} and \eg][]{veverka2001b}. These features, referred to as ``ponds", are characterised by smooth, level surfaces that are sharply delineated \citep{robinson2001, cheng2002}. They are found preferentially at low latitudes and in the bottom of small ($<$1 km) craters or other topographic lows \citep{robinson2001,cheng2002}, however, this may be due to observational biases \citep{roberts2014}. The bottoms of the ponds are often offset in a direction towards the downslope of the crater \citep{veverka2001} and recent results have found that the pond floors are not as flat as originally believed \citep{roberts2014b}.

\paragraph{Linear features on Eros - }

On the surface of Eros several lineations can be observed including chains of craters, sinuous and linear elongated depressions and topographic ridges \citep{veverka2000}. 
Such lineations are similar to those observed on the Martian satellite, Phobos \citep{thomas1979}. \cite{prockter2002} explain that, on Eros, these linear features, or grooves, exist on a global scale (prominent wide troughs and ridges several kilometres in length), a regional scale (chains of craters and straight-edged grooves several hundreds of meters long) and also on very local scales (closely spaced ridge and trough terrains tens of meter in scale). The large variations in directions, patterns and relative ages of the lineations indicate that they were formed during many different and unrelated events \citep[see \sect{Processes} and Marchi et al. (this volume);][]{veverka2000, prockter2002, thomas2002, robinson2002}. For a full map and analyses of the linear features on Eros see \cite{buczkowski2008}.

\paragraph{Depth and character of Eros' regolith - } 
Eros has a widespread unconsolidated regolith of depths that are typically several tens of meters in thickness, but not uniform over the surface \citep{cheng2002b, barnouin2001,veverka2001b,robinson2002}.  The heterogeneity of the regolith depth distribution is probably caused partially by the asymmetric nature of crater ejecta blankets  \citep[a consequence of the asteroid's rotation, see \sect{Processes} and][]{geissler1996} and is further accentuated by the irregular spacing of craters and the subsequent downslope motion and regolith transport that, as discussed above, appears to occur commonly on the surface of Eros \citep{robinson2002}.

The surface of Eros is extremely rough and the surface roughness is approximately self-affine from scales of a few meters to hundred of meters \citep{cheng2002b}. The regolith particles range in size from the fine ($\ll$cm - sized) dust particles found in the ponds to the numerous ($>$10$^4$) large ($>$10 m) ejecta blocks of boulders at the extreme large end of the particle size distribution \citep{thomas2002}. The morphology of these blocks ranges from angular to fractured to disaggregated \citep{robinson2002} and their size distribution is described adequately by a power law with a slope of about -3 on a cumulative plot (\fig{Cumulative}). For more information about the nature of these boulders see Marchi et al. (this volume).

\begin{table*}[t]
\footnotesize
  \centering
\caption{Characteristics of the asteroids discussed in detail in this chapter. $^{(1)}$\cite{cheng1997}; $^{(2)}$\cite{yeomans2000}; $^{(3)}$\cite{miller2002}; $^{(4)}$\cite{veverka2000};$^{(5)}$\cite{fujiwara2006}; $^{(6)}$\cite{abe2006}; $^{(7)}$\cite{scheeres2006b}; $^{(8)}$\cite{schulz2012}; $^{(9)}$\cite{sierks2011}; $^{(10)}$\cite{lamy2010}; $^{(11)}$\cite{thomas2012}; $^{(12)}$\cite{russell2011b};$^{(13)}$\cite{russell2012}}
  \begin{tabular}{l c c c c c c}
  \\
  \hline
  \hline
  \bf{Asteroid} & \bf{Space Mission} & \bf{Mean diameter} &  \bf{Bulk density} & \bf{Rotation period} & \bf{Surface acceleration} & \bf{Escape speed} \\  &  & \bf{(km)} &  \bf{(g cm$^{-3}$)} & \bf{(h)} & \bf{(cm s$^{-2}$)} & \bf{(m s$^{-1}$)} \\
  \hline
(433) Eros & NASA NEAR$^{(1)}$ & $\sim$17$^{(2)}$ & 2.7$^{(2)}$ & 5.3$^{(3)}$ & 0.23 - 0.56$^{(3)}$ & $\sim1^{(4)}$ \\
 (25143) Itokawa & JAXA Hayabusa$^{(5)}$ & $\sim 0.32^{(5)}$ & 1.9 $^{(6)}$ & 12.1$^{(7)}$ & 2.4e-3 - 8.6e-3$^{(7)}$ & 0.1-0.2$^{(5)}$ \\
 (21) Lutetia & ESA Rosetta$^{(8)}$ & $\sim$99$^{(9)}$ & 3.4$^{(9)}$ & 8.2$^{(10)}$ & $\sim$5$^{(11)}$ & $\sim$70$^{(11)}$ \\
 (4) Vesta & NASA Dawn$^{(12)}$ & $\sim$526$^{(13)}$ & 3.5$^{(13)}$ & 5.3$^{(13)}$ &  $\sim$25 & $\sim$363  \\
 \hline
  \end{tabular}
\label{t:asteroids}
\end{table*}

\subsection{Asteroid (25143) Itokawa}\label{s:Itokawa}

Compared to Eros, the NEA (25143) Itokawa (hereafter simply Itokawa; \fig{Asteroids}, \tbl{asteroids}) was found, astonishingly, to have entirely different structural and surface properties despite their similar taxonomic class. The reason for these different properties is not clearly understood, but perhaps this shouldn't have been surprising; because of their size (mass) difference, if gravity is the discriminator, then Itokawa is expected to be as different from Eros, geologically, as Eros is from the Moon \citep{asphaug2009}.

 One of the most remarkable features of Itokawa is the global shape, which seems to consist of two parts: a small ``head" and a large ``body" separated by a constricted ``neck" region \citep[\fig{Asteroids} and][]{fujiwara2006,demura2006}. { It is highly likely that Itokawa is a rubble pile asteroid rather than a monolithic body \citep{fujiwara2006}.   The low bulk density of Itokawa (\tbl{asteroids}) provides further evidence for the rubble pile interior structure with estimates suggesting that Itokawa's macroporosity may be as high as $\sim$41\% \citep{fujiwara2006}. However, these density measurements do not rule out the presence of a core on the order of 100 m in size.  }
 
\paragraph{Depth, character and migration of Itokawa's regolith - } 

Two different types of terrain - rough and smooth - are observed on Itokowa \citep{saito2006}.   The rough deposits consist of numerous boulders \citep{fujiwara2006} and typically exhibit variations in elevation that range from 2-4 m over small lateral distances \citep{barnouin2008}. The very highest and roughest parts of the asteroid are covered in large gravel and boulders and are completely devoid of all particles smaller than 1 cm in size \citep{barnouin2008}.   The smooth terrains - Muses Sea and Sagamihara - coincide with the low-gravitational potentials and are generally homogeneous, featureless and relatively flat (slopes $<$8$^\circ$). This is consistent with a loose granular layer that has been allowed to seek out its minimum energy configuration after the formation of the asteroid \citep{miyamoto2007, fujiwara2006,yano2006,riner2008}.  This idea is further reinforced by the close-up images and measurements taken during the touch down of the Hayabusa spacecraft; these indicate that small regolith particles are being transported into the Muses Sea region and are gradually covering up the boulder-rich surface \citep{miyamoto2007,barnouin2008}. The regolith depth in the smooth regions on Itokawa is estimated to be approximately 2.5 m \citep{barnouin2008,cheng2007}.

In general Itokawa's regolith appears to be dominated by grains $>$1 mm in size \citep{miyamoto2007}.  That said, the regolith particles that were returned to Earth are fine-grained \citep[size range between 3-180 $\mu$m, but most { $<$ 10} $\mu$m; ][]{nakamura2011}. The apparent absence, or at least the small quantity, of fines on the surface of Itokawa may be explained by processes such as electrostatic levitation combined with solar radiation pressure \citep{lee1996, scheeres2005}, segregation of the fines towards the interior of the body \citep{asphaug2007, miyamoto2007} or simple higher ejection velocities following impacts making reaccumulation difficult \citep{nakamura1994}. Some of these processes will be discussed later in \secttwo{Environment}{Processes}. 

\begin{figure}[h!]
\begin{center}
\includegraphics[scale=0.28]{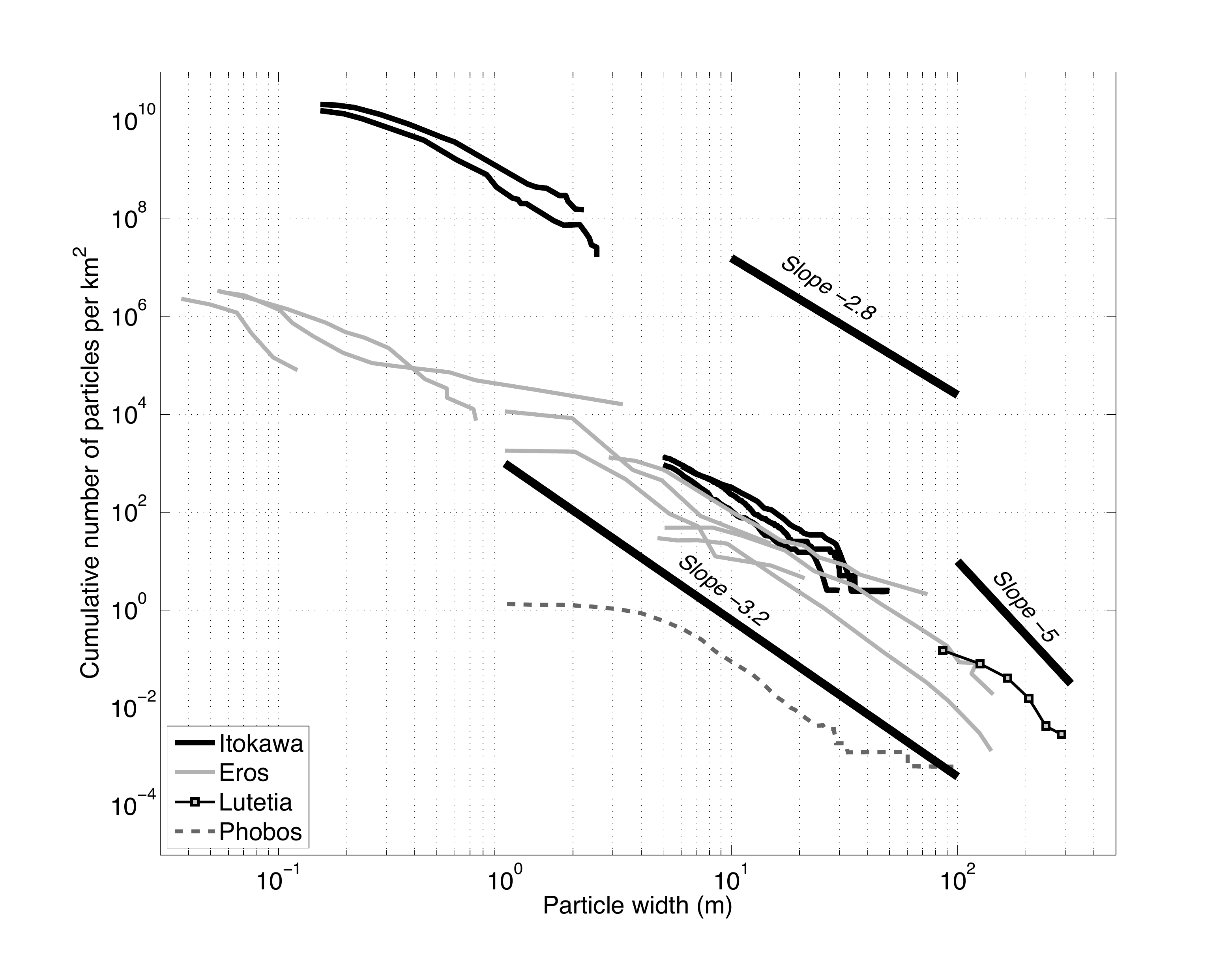} 
\caption{Measured cumulative size distribution as a function of particle size on the surfaces of asteroids and Phobos. Data combined from several papers. Itokawa: \cite{miyamoto2007}, \cite{saito2006}, \cite{michikami2008}, \cite{mazrouei2012}; Eros: \cite{thomas2001}; Lutetia: \cite{kuppers2012}; Phobos (a Martian satellite with a mean diameter of $\sim$22 km): \cite{thomas2002}. For the papers in which the cumulative number of particles was given, this has been approximately converted to cumulative number per square kilometre using the information provided in the respective papers.  A shallower slope { may} indicate that boulders have experienced less processing, including breaking, sorting and transporting \citep{thomas2002}. }
\label{fig:Cumulative}
\end{center}
\end{figure}

The size distribution of boulders on Itokawa's surface is estimated to be a power law with a slope of -2.8 to -3.0 on a cumulative plot (\fig{Cumulative}).  { It is possible, however, that the observed distributions on Itokawa may be related to the preferential displacement of some block sizes relative to others, and the settling locations of differing sized blocks.}  The abundance of meter-sized boulders \citep[particularly on the western side;][]{fujiwara2006}, and the fact that decameter-sized boulders exist \citep[the { length of the} largest boulder is approximately one tenth of the { length} of Itokawa itself;][]{saito2006}, indicate that they may have been produced during a catastrophic disruption event, consistent with the rubble pile structure \citep{fujiwara2006, michel2001}.

\paragraph{Further evidence for an active regolith on Itokawa - }

At the boundary of the Muses Sea region with the rough terrain, boulders are typically piled on top of each other without being buried by fines. The larger sized gravels tend to lie over the smaller particles and are aligned with directions coincident with the local gravity slope \citep{miyamoto2007}. This type of organisation of gravels is referred to as imbrications, in this case with the longest axes of the gravel being preferentially orientated transverse to the granular flow. The positions and orientations of all of the particles indicate that they are stable against local gravity and that the migrations were gravity-induced \citep{miyamoto2007}.  

\begin{figure}[htbp]
\begin{center}
\includegraphics[scale=0.18]{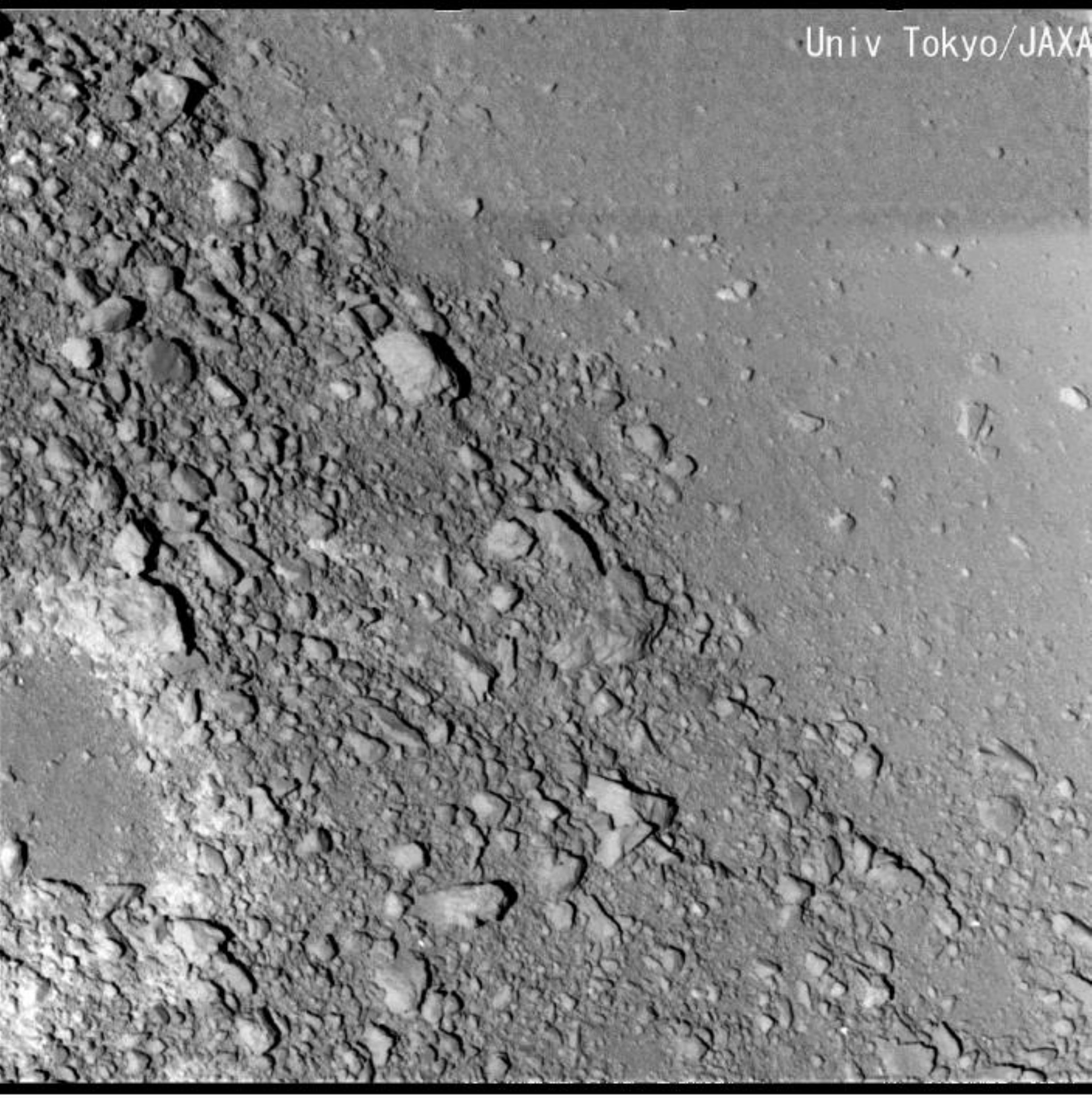} 
\includegraphics[scale=0.28]{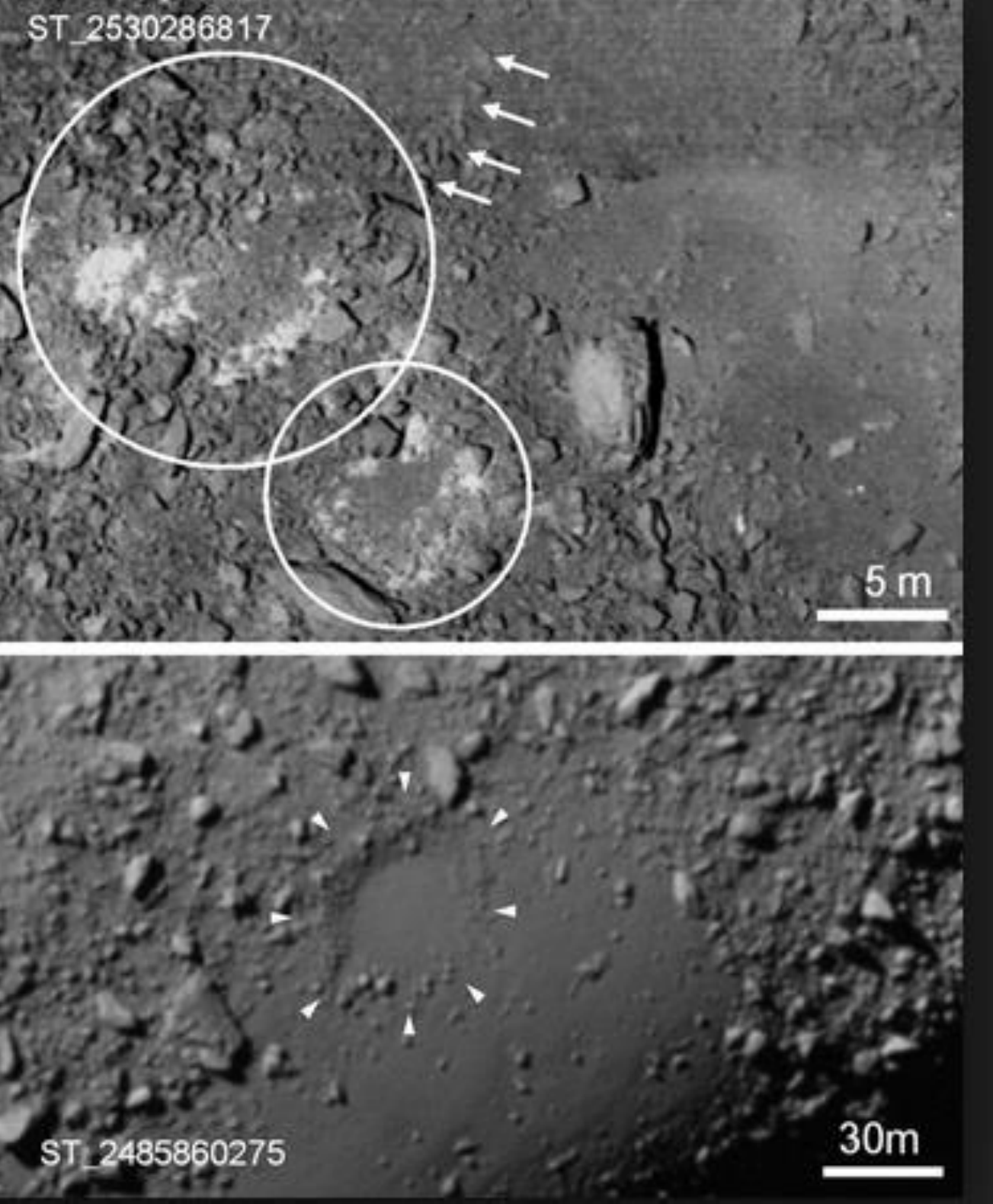}
\caption{Evidence of regolith transport on Itokawa - LEFT: High resolution image of the boundary area between Muses Sea and the rough terrains. Piles of gravel can be seen on the uphill sides of boulders. Such a characteristic is similar to terrestrial landslide; TOP RIGHT: Image of the Neck area of Itokawa (13cm/pixel) showing piles of angular boulders at the lower part of the image. Crater-like depressions are shown by the circles and the arrows indicate the debris, which appears to have drained from the rim of the upper crater towards the smooth terrain; BOTTOM RIGHT: A circular depression that appears to be filled with finer particles.}
\label{fig:ItokawaRegolithTransport}
\end{center}
\end{figure}

Evidence of landslide-like deposits can be seen in \fig{ItokawaRegolithTransport}. There are large boulders that have blocked the migration of smaller particles, resulting in piles of smaller particles on the uphill sides of the boulders \citep{miyamoto2007}.   Unlike the surface of Eros, Itokawa is very heterogenous in colour and albedo, with brighter surfaces being found in three main areas: areas with steeper slopes, areas of local high terrain and apparently eroded areas, \eg crater rims \citep{saito2006}.  \cite{saito2006} suggest that this dichotomy is due to dark surfaces being removed, leaving the fresh regolith newly exposed at the surface, as observed on Eros.

All of these observations give a strong indication that regolith on the surface of Itokawa has been relocated since the initial accumulation or deposition.

\paragraph{Craters and crater morphology on Itokawa - }

Itokawa has very few craters in general (the total number of craters on Itokawa is $<$100 over the entire surface including indefinite candidates) and absolutely no distinct craters $<$1 m in diameter \citep{saito2006, fujiwara2006}.  Those craters that do exist on the rough and transitional terrains, and that retain their regolith, are filled with finer particles, similar to the ``ponds'' seen on Eros.  The best example of a crater on the surface of Eros - Komaba -  is located near the edge of the highlands. It has a small depth-to-diameter ratio (0.09) consistent with crater formation in a coarse granular target, a flat floor and is surrounded by brighter rims \citep{saito2006, barnouin2008}.   In addition, there appears, unusually, to be no apparent correlation between the locations of boulders and craters \citep{michikami2008}. These observations are further evidence of regolith motion and suggest that a mechanism is filling in and erasing the craters on the surface of Itokawa. Such a mechanism may well be seismic shaking, as proposed to explain paucity of small craters on Eros (see \sect{Processes}). Alternatively, it is also possible that Itokawa could have been generated relatively recently in the main belt before being moved to it's current orbit \citep{saito2006}.

\paragraph{Linear features on Itokawa - }

Unlike Eros, there are no global lineaments on Itokawa \citep{fujiwara2006}.  However, on the body of Itokawa subtle local linear features can be observed. These features, caused by the alignment of boulders \citep{cheng2007}, are not as tall as other structures such as large boulders. Nonetheless, they are an important contributor to the topography of Itokawa due to their large lateral extent \citep{barnouin2008}. 

\subsection{Asteroid (21) Lutetia}\label{s:Lutetia}

The main-belt asteroid (21) Lutetia (hereafter simply Lutetia; \fig{Asteroids}, \tbl{asteroids}) has a highly complex surface geology with significant interactions between ancient and more recent structures \citep{sierks2011,thomas2012}. 

\paragraph{ Craters and crater morphology on Lutetia - }

The higher gravity and escape velocity on Lutetia have provided an environment for continuous ejecta patterns with obvious relations to the impact from which they formed \citep{massironi2012}. The typical depth-to-diameter ratio of craters on Lutetia is 0.12 but values have been observed ranging from 0.05 to 0.3 \citep{vincent2012}. The distribution of depth-to-diameter ratios varies depending on the region of Lutetia's surface indicating that, not only are there variations of physical properties across the surface, but there are also differences in the surface evolutionary processes \citep{vincent2012}.

\cite{thomas2012} divide the craters on Lutetia into four different categories: standard craters, buried or partially filled craters, distorted or cut craters that have been disturbed by lineament formation, and morphologically non-standard impact structures. By the latter they refer to craters that are not typically bowl-shaped and/or do not have a round rim and the strange form is not obviously linked to linear features. Such unusually shaped craters \citep[for examples see][]{thomas2012} could be the result of oblique impacts \citep{thomas2012, herrick2006,krohn2014} but other mechanisms have also been proposed \citep{vincent2012}.

Similarly to the surface of Eros, Lutetia's surface exhibits a paucity of small ($<$1 km) craters. This could perhaps be explained by seismic shaking; however, there is also a depletion in craters of sizes up to 8 km, which is more difficult to attribute to seismic shaking. The craters that have been deformed by linear features are additional evidence that the surface has been modified since the crater formation \citep{sierks2011}.

\paragraph{Depth and character of Lutetia's regolith - } 

The surface of Lutetia is covered by an extensive regolith, similar to that of the Moon \citep{coradini2011}. Nonetheless, Lutetia's surface is very heterogeneous.  Images taken during the close approach of ESA's Rosetta spacecraft allowed Lutetia's surface to be separated into several distinct regions \citep[for a detailed map of the Lutetia regions see][]{sierks2011, massironi2012}.  Some regions are very old and heavily cratered with significant deformation by linear features, while others exhibit sharp morphological boundaries. The Baetica (North Pole) region contains a cluster of craters, created from a series of superposed impacts \citep{massironi2012} this is one of the most prominent features imaged on Lutetia's surface. The extremely low crater density and lack of linear features in this region can perhaps be attributed to the covering of smooth regolith material, probably the ejecta blanket from the crater cluster \citep{sierks2011,vincent2012}.
 
Lutetia's regolith is estimated to be up to $\sim$600 m in depth \citep{vincent2012}. {This estimate is based on the thickness of the ejecta blanket of the largest crater assuming a uniform gravity field and may, therefore, be improved with a more detailed study of regional ejecta geophysics taking into account the complex gravitational field of Lutetia. }Surface slopes can exceed 30$^\circ$ in some places but are generally less than this \citep{sierks2011,thomas2012}. The size distribution of blocks on Lutetia is reported to be a steep power-law of -5 \citep[\fig{Cumulative} and][]{kuppers2012}. { It is noted, however, that the method used by \cite{kuppers2012} for binning the boulders is different to the method used by other research groups. }

\paragraph{Evidence for an active regolith on Lutetia - }

Diverse evidence for regolith motion was observed inside the large crater cluster in the North pole (Baetica) region. The observations include albedo variations with bright regions on the steep slopes indicative of relatively recent landslides (as observed inside craters on Eros and on Itokawa), deposits of smooth and fine particles with boulders, and apparent landslide deposits \citep{sierks2011}. In addition, observations show craters that have poorly defined rims as a consequence of multiple landslides \citep[\eg Fig. 5 of ][]{thomas2012}. Rocky outcrops are also visible at what appears to be the source of the landslides \citep{thomas2012}.

\paragraph{Linear features on Lutetia - }

\begin{figure}[htbp]
\begin{center}
\includegraphics[scale=0.42]{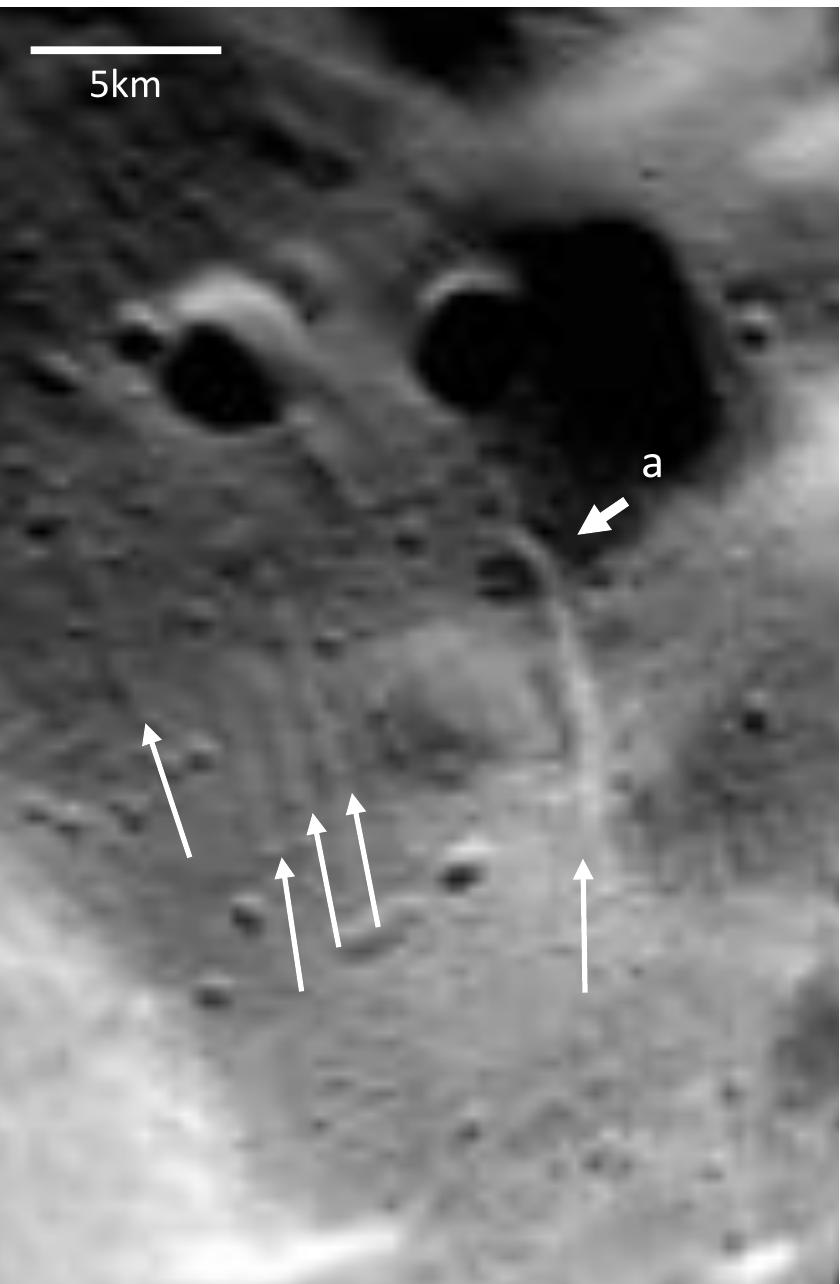} 
\includegraphics[scale=0.42]{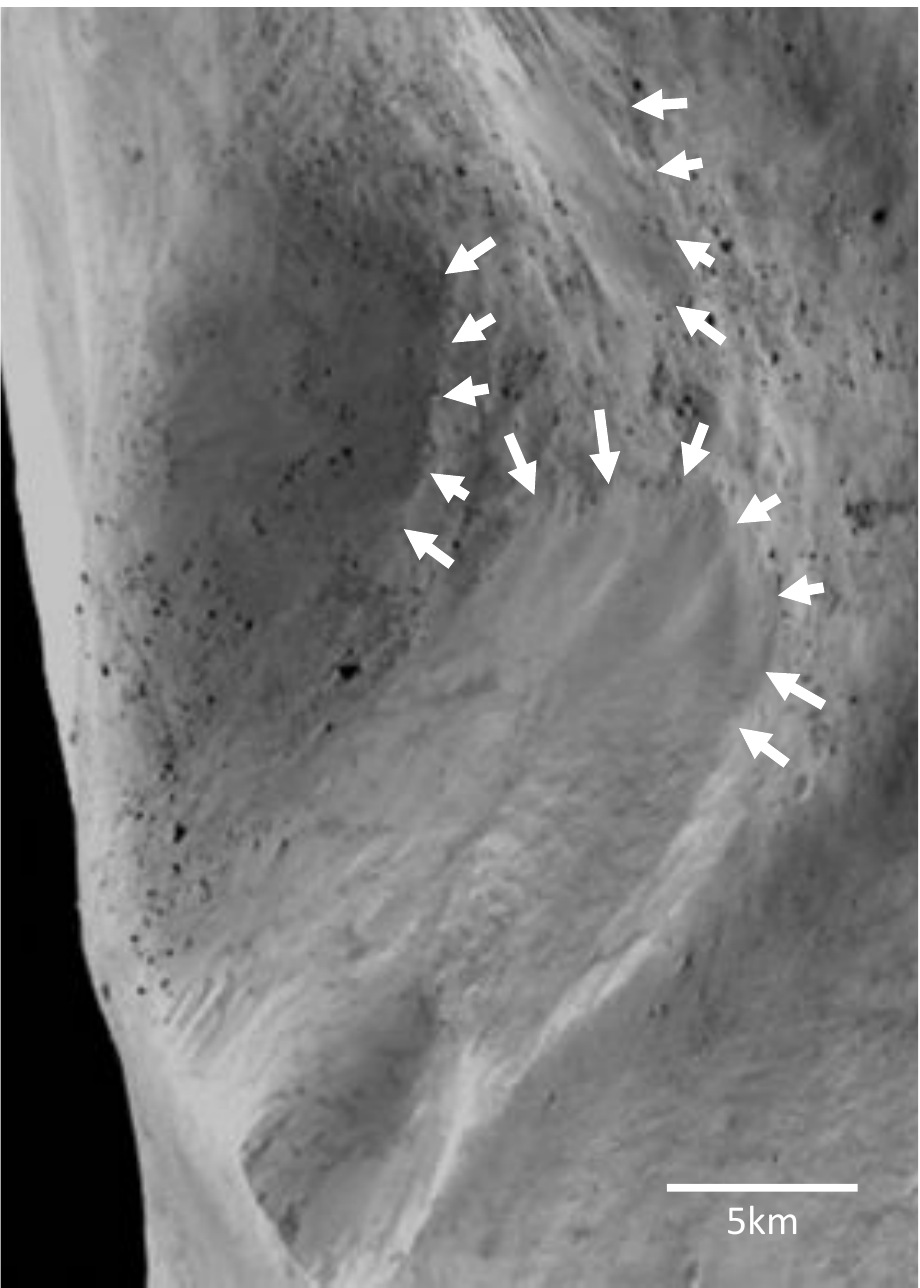}
\caption{Closeup images of Lutetia - LEFT: Curvilinear features. Some linear features cut crater rims (a) implying these features postdate the craters (ID 224598, ESA 2010 MPS for OSIRIS Team MPS/UPD/LAM/IAA/RSSD/INTA/UPM/DASP/IDA); RIGHT: The central crater cluster in Baetica showing landslides (arrows) and numerous boulders (ID 216820, ESA 2010 MPS for OSIRIS Team MPS/UPD/LAM/IAA/RSSD/INTA/UPM/DASP/IDA)}
\label{fig:LutetiaFaultsGrooves}
\end{center}
\end{figure}

Lutetia displays a huge number of lineaments (\eg \fig{LutetiaFaultsGrooves}) that can be found over the entire imaged surface, with the exception of two young regions \citep{thomas2012}. {The orientation of these linear features, which are similar in appearance to those on Eros discussed in \sect{Eros}, has been linked to three impact craters \citep{besse2014}. } The linear structures have been classified into several types by \cite{thomas2012}: irregular troughs, large faults and tectonic troughs, organised linear reflectance variations and narrow faults, rows of coalesced pits (known as pit-chains), intra-crater trenches, intra-crater layers and ejecta layers and, finally, scarps and ridges. The most striking linear feature on Lutetia's surface is the very long ($\sim$10 km) and wide ($\sim$1.2 km maximum width) groove in the Noricum region \citep{thomas2012}. This groove is situated on a local topographic high and is approximately 100 m in depth \citep{sierks2011}.  For a very complete discussion of the lineaments on Lutetia, including multiple examples, see \cite{thomas2012}. 

\subsection{Asteroid (4) Vesta}

(4) Vesta (hereafter simple Vesta; \fig{Asteroids}, \tbl{asteroids}) is the second most massive main-belt asteroid and is one of the fastest rotators of the large asteroids. Vesta's surface has a complex topography at all spatial scales \citep{jaumann2012}. One of the most dramatic discoveries on the surface of Vesta is an 18 km high mountain in the centre of a huge (460 km-wide) crater named the Rheasilvia basin.  This peak is the second highest in the Solar System after Olympus Mons on Mars. 
 
\paragraph{Depth and character of Vesta's regolith - } 

\begin{figure}[hbp]
\begin{center}
\includegraphics[scale=0.6]{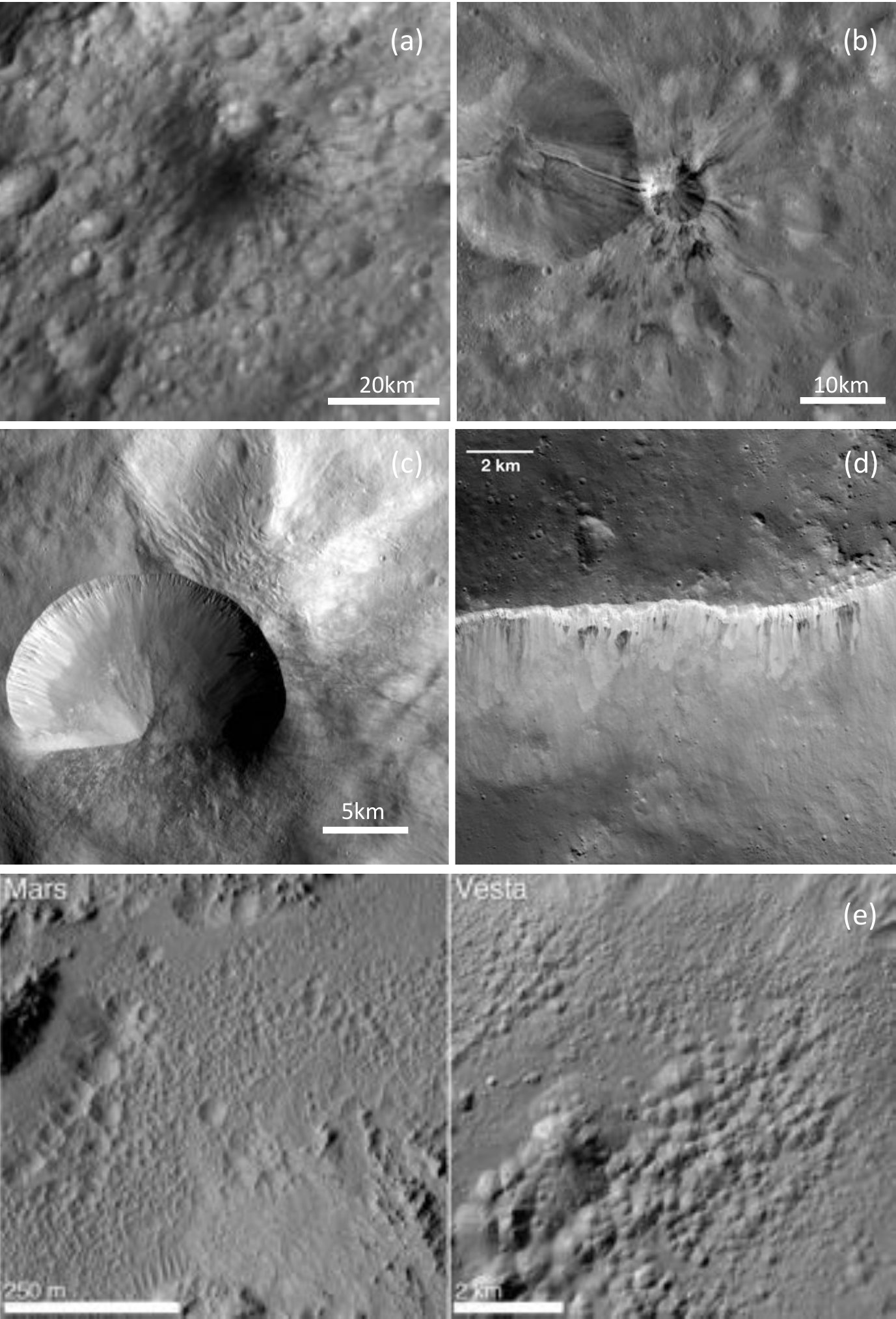} 
\caption{Surface processes on Vesta; (a) Dark hill (PIA14689, modified), (b) Fresh crater (center) with bright and dark rays (PIA15045, modified), (c) Crater on a slope with a sharp crest uphill and slumping material covering the lower rim PIA15495, (d) Dark and bright material at the rim of Marcia crater (NASA/JPL-Caltech/UCLA/MPS/DLR/IDA/LPI/ASU), (e) Pitted terrains on Mars (left) and Vesta (right) (PIA16185, modified).}
\label{fig:VestaSurface}
\end{center}
\end{figure}

The thickness of Vesta's regolith is estimated to be approximately 800 m \citep{jaumann2012}. The surface slopes on Vesta can exceed 40$^\circ$ and there are a considerable number of steep slopes that may be indicative of intact bedrock beneath or the presence of cohesive forces in the regolith. Dark materials, of likely exogenic origin (carbon-rich low-speed impactors), are distributed unevenly across Vesta's surface \citep[\fig{VestaSurface}(a);][]{jaumann2014}.

\paragraph{Craters and crater morphology on Vesta - }

Craters on Vesta display a wide range of degradation states from fresh craters with unmodified rims to impact crater ruins showing almost no visible rims \citep{jaumann2012}. Depth-to-diameter ratios are similar to Lutetia, varying from 0.05 to 0.4 with a mean of 0.17 \citep{jaumann2012,vincent2012,vincent2013}. The northern hemisphere is observed to be heavily cratered whereas the southern hemisphere shows comparatively fewer craters, most probably due to the relatively recent basin forming impacts near the south pole \citep{vincent2013}. Shallower craters are found in the oldest regions on the surface of Vesta, as would be expected due to progressive crater degradation. The deep, loose regolith in the younger southern hemisphere may also aid the formation of deeper craters \citep{vincent2013}.

{ Topography plays a much more important role in crater formation and evolution on small bodies and moons than on terrestrial planets. For example, Vesta's ratio of observed relief to size \citep[15\%;][]{williams2013} is significantly greater than for terrestrial planets \citep[1\%;][]{jaumann2012}. } Strongly asymmetric craters have been seen on the many steep surfaces of Vesta (\fig{VestaSurface}(c)). During impacts on steep slopes ejecta is prevented from being deposited in the uphill direction and slumping material superimposes the deposit of ejecta on the downhill side \citep{krohn2014}. This leads to craters with a smoothed downslope rim that is often covered by the asymmetric ejecta \citep{jaumann2012}.   Resurfacing due to impacts, gravitational modifications and seismic shaking are important geophysical processes that not only add to the complexity of Vesta's surface evolution, but also substantially alter Vesta's morphology \citep{jaumann2012}. Young bright and dark-rayed craters and their ejecta field are superposed across the surface of Vesta \citep[\fig{VestaSurface}(b);][]{williams2014, yingst2014}. { Pond-like deposits are also seen on the surface of Vesta. Similarly to on Eros, they tend to have a downslope asymmetry within craters on slopes, and show no evidence for regolith flows into the craters and depressions \citep{jaumann2012, cheng2002}. However, given the more important gravity on Vesta (compared to Eros) and the larger size of such pond-like features, their formation may simply be due to standard crater slumping, rather than a process by which external material is transported into the crater.  }

\paragraph{Evidence for an active regolith on Vesta - }

As on the surfaces of the other asteroids discussed so far, extensive evidence of regolith mobility has been observed on Vesta; slumps of material, scarps beginning at the top of a slope, dark and bright material emanating from the rims or walls of impact craters, or running downslope into the crater bowl \citep[\fig{VestaSurface}(d);][]{jaumann2012, williams2013,yingst2014}. Lobate, flow-like features are generally observed in close proximity to impact craters or in steep slopes \citep{williams2013}. These features are interpreted as gravity-driven mass flow deposits, impact ejecta deposits or, for a small number of features, impact melt deposits \citep{jaumann2012, williams2013}.

On Vesta, the seismic shaking created by the giant basin forming impacts probably contributed to smoothing and erasure of small features well beyond the extent of the ejecta blankets \citep{vincent2013}.  Mixing of regolith materials (\eg the dark exogenic materials with impact ejecta) is evident on Vesta's surface \citep{jaumann2014,pieters2013}. There is a also dearth of large-scale volcanic features on the surface of Vesta, compared to what was expected \citep[\eg][]{wilson1996,mcsween2011}. \cite{jaumann2012} suggest that the lack of such features may be due to extensive cratering, regolith formation and resurfacing that has removed the evidence of large-scale volcanism that ceased early in Vesta's history \citep{williams2013, jaumann2012}. This will be discussed further in \sect{Processes}.

\paragraph{Linear features on Vesta - }

Large equatorial and northern troughs appear on Vesta's surface (\fig{Asteroids}). The equatorial troughs are wide, flat-floored and bounded by steep scarps along $\sim240^\circ$ of longitude, while in the remaining longitude muted troughs, grooves and pit crater chains are evident \citep{jaumann2012}. The northern troughs display gentler slopes, rounded edges and considerable infilling. This, combined with the heavy cratering, suggests that they are much older than the equatorial troughs \citep{jaumann2012}. The centre positions of these circular troughs correspond to the centre of Vesta's two southern basins indicating that the formation of the troughs and the basins are very likely related \citep[see Fig. 2 of ][]{jaumann2012}.

\section{\textbf{AN INTRODUCTION TO GRANULAR MEDIA}}\label{s:Grains}

Granular materials are unlike solids, in that they can conform to the shape of the vessel containing them, thereby exhibiting fluid-like characteristics. On the other hand, they cannot be considered a fluid, as they can be heaped \citep{gudhe1994}. The study of granular dynamics is incredibly complex and constitutes an entire field of research by itself.  In fact, P. G. De Gennes, a French physicist and Nobel Prize laureate, said that, \emph{``For physicists, granular matter is a new type of condensed matter; as fundamental as liquid, or solid; and showing in fact two states: one liquid-like, one solid-like. But, there is yet no consensus on the description of these two states. Granular matter, in 1998, is at the level of solid state physics in 1930."} \citep{deGennes1999}.  {  This is not to say that granular matter has not been studied; ancient Egyptians did indeed know how to work with it, at least at an empirical level \citep{fall2014}, Ernst Chladni \citep{chladni1787} and Michael Faraday \citep{faraday1831} studied the interaction of grains and fluids, and Geosciences have also long dealt with its complexities on the surface of the Earth.}

On Earth we can observe granular materials involved in dramatic avalanches and rockslides, as well as active sand dunes moving across deserts. Industries also handle several different types of granular materials. Some examples are tablets or powders in the pharmaceutical trades as well as agricultural products such as wheat, oats, rice and other cereals and sands in the construction industry. Theoretical models of granular dynamics are also widely employed to understand traffic flow and even crowd dynamics.

\paragraph{What is a granular material?} 

The term granular material is most often used to describe a material containing a large number of particles that interact with each other through dissipative contact forces \citep{richard2005, jaeger1996}.  In these aggregates, though each individual grain can be adequately described by Newtonian physics, a collection of grains offers complex behaviour which is often extremely sensitive to their external conditions (such as external forcing). A granular material is a material for which the relevant energy scale is the potential energy rather than the thermal energy \ie particles in a granular material are massive enough for their potential energy to be orders of magnitude larger than their thermal energy \citep{schroter2005}.  For example, a typical grain of sand of mass $m$, raised by its own diameter $d$, in the Earth's gravity $g$, will have potential energy $mgd$ which is at least 10$^{12}$ times the thermal energy $k_BT$ at room temperature on Earth \citep{jaeger1996}.  

The size of the constituent particles is closely linked to the type of interactions between the particles that will dominate the behaviour of the aggregate.  On Earth the approximate size at which dissipative contact interactions dominate is 100 $\mu$m; at grain sizes $<$100 $\mu$m, humidity and van der Waals forces will influence the particle interactions.  Additionally, if present, the interstitial fluid will also influence the dynamics of the grains depending on the density of the fluid and the grain size \citep{burtally2002, biswas2003}.  We will discuss in \sect{Environment} how the importance of some of these forces changes in the low-gravity environment of an asteroid.

\paragraph{Basic characteristics of granular materials - } 

Granular materials exhibit several characteristics that make them interesting but equally very difficult to model and understand.  From the definitions presented above, some characteristics can be extracted, whilst others come from observation:  (1) the grains that form a granular material are solid; (2) grain-grain interactions are highly dissipative; (3) potential energy, more than temperature, of the system is the relevant parameter; (4) granular materials are thixotropic; this means that they exhibit solid-, liquid- and gas-like behaviour: (5) friction, globally understood to be a combination of surface-surface friction and geometrical interlocking that prevents motion, makes aggregates able to sustain shear stress and contribute to the dissipative nature of grain-grain interactions.

\begin{figure}[ht!]
\begin{center}
\includegraphics[height=1.4in]{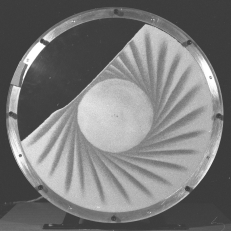} 
\includegraphics[height=1.4in]{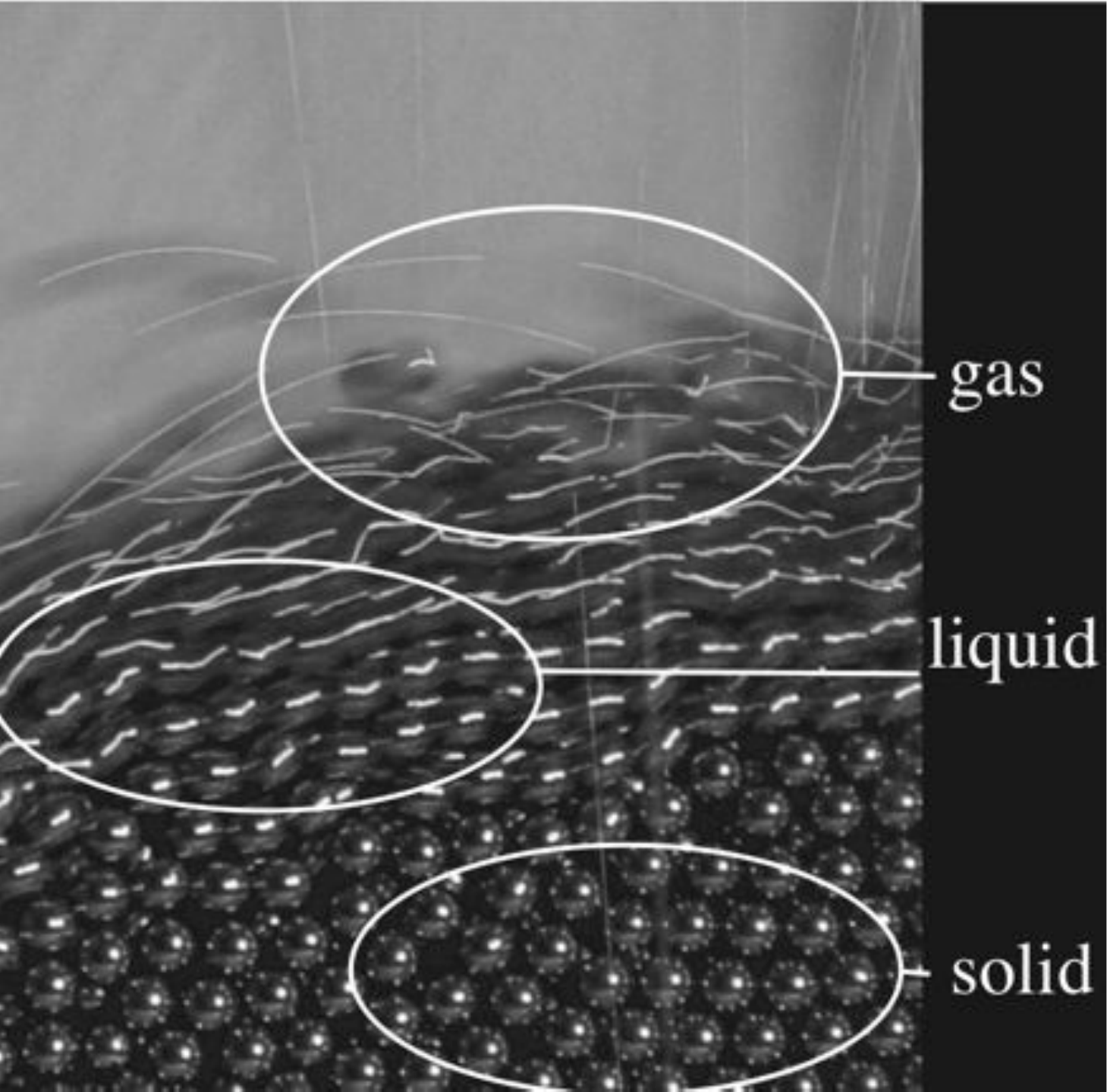}
\includegraphics[height=1.4in]{Murdoch_Fig7c.pdf} 
\caption{{Top Left: Segregation of particles in a tumbler -} The large sugar crystals (white) and small iron particles (black) segregate in the tumbler as described in \cite{gray2006}. Credit: N. Gray, University of Manchester; {Top Right: Solid, liquid or gas? -} An illustration of the solid (bottom), liquid (middle), and gas (top) flow regimes obtained by pouring steel beads on a pile. Credit: O. Pouliquen and Y. Forterre; {Bottom: Force chains in a granular material} - Photoelastic image of a system that has been jammed by applying simple shear strain from an initially force-free state.  The apparatus used here applies shear from the boundaries and also from the base, which consists of individual slats which deform affinely with the boundary. Credit: J. Ren and R. P. Behringer, Duke University.}
\label{fig:granularpics}
\end{center}
\end{figure}

From these basic characteristics, some phenomena result. For example, \fig{granularpics} (upper right) shows the solid, liquid, and gas flow regimes obtained in an avalanche-like situation. In a solid-like state, such as a heap or pile, the material is said to be ``quasi-static" as the individual particles are in a stable mechanical equilibrium with their local neighbours. In the liquid-like and gas-like states the material is said to ``flow". Dense flows (liquid-like state) are dominated by many-body interactions and occur when particles have long-lived contacts with many neighbours. In rapid or dilute flows (the gas-like state) there are no enduring contacts and the collision time is much smaller than the time between collisions {(\citealt{andreotti2013} provide a thorough exposition of these topics).}

Flowing granular materials can segregate according to particle properties (size, density, shape, and more) making granular media highly heterogeneous (\fig{granularpics} [upper left]). The phenomenon of segregation is a continued source of frustration for industries \citep{mccarthy2009}, however, segregation may help us to explain several geological features observed on the surface of asteroids and discussed at the beginning of this chapter.

The final, but very important, property of granular materials is the non-linear transmission of force between particles via force chains.  A force acting on a granular material is distributed through a complex force distribution network that depends on the positioning and packing of the individual particles. This grain network resists reorganisation when stressed and imposes a granular drag force when a solid object is pushed through the material \citep{constantino2008}. \Fig{granularpics} (lower) shows the force chains inside a granular material.  The presence of force chains can induce complex stresses at the sides of grain silos \citep{schwartz2012}, preventing explosions at the bottom and instead leading to ruptures at the sides \citep{janssen1895,jaeger1996}. This can be linked to the non-local effects granular materials exhibit \citep{nichol2010} and may even cause asteroids to feel long range consequences of small events such as meteoroid impacts \citep{murdoch2013b}.

\paragraph{{Theoretical Frameworks - }} 

 As shown in the paragraphs above, granular matter can present solid-, liquid-, and gas-like behaviours, all at the same time.  Although a complete theory to describe each behaviour simultaneously has yet to be put forward (\eg \citep{jop2006, midi2004}), different regimes (or states) can be modelled within certain frameworks. Depending on which of these theoretical frameworks is best-suited to the regime at hand, the properties of the grains will be described by different sets of parameters, \eg friction, elastic moduli, viscosity, and restitution coefficients among others.

The elasticity-perfect plasticity models can be used for the static case (\eg \citealp{holsapple2004, holsapple2006}). These models belong to the field of Continuum Mechanics and, as the name would suggest, they treat a granular media as continuous.  This can be done under one assumption, the size of the grains that form the media (or soil) are very small compared with the typical length scale or the size of the sample.  The dynamics of the media is modelled through yield criteria such as Mohr-Coulomb or Drucker-Prager in which the main parameters are angle of friction and cohesive strength.  The pressure and shear stress (both derived from the principal stresses of the stress tensor) define the stress state of the media and are average quantities that in reality result from the contacts between particles and the interactions between their surfaces.

Fluid mechanics equations are used for dense flows, or for when grains begin to flow like a liquid (\eg \citealp{haff1983,forterre2008}).  Within this framework, the dynamics of the medium is described through a continuity equation, derived from the conservation of mass principle; a momentum equation, derived from the conservation of momentum principle (in the case of granular materials, a description of viscosity must be included); an energy equation, derived from the conservation of energy principle; and an equation of state relating the three conservation equations.
 
Kinetic Theory (\eg \citealp{jenkins2002,brilliantov2010}), used for dilute, highly dynamical (gas-like) systems makes the following assumption: the particles only have binary collisions.  Of course, in a real system this is not the case as multi-particle collisions may also occur; however, they are determined to be too rare to be taken into account.  The validity of the assumption is related to the density of the granular system (must be safely below jamming density) and the duration of the collisions (must be short in order to avoid the occurrence of three or more particles in simultaneous contact).  Thus, the particles of this system are idealised as hard spheres (instantaneous collisions).  In this particular regime, the concept of granular temperature can be defined.  A ``granular temperature" can be defined in multiple ways, but essentially it is some measure of the average of energy fluctuations exhibited by a collection of grains.  An example of one such definition, for a collection of $N$ grains of average velocity $\bvec{\bar{v}}$ and average spin $\bsym{\bar{\omega}}$, with each grain having a mass $m_i$, velocity $\bvec{v}_i$, moment of inertia $I_i$, and spin $\bsym{\omega}_i$, the granular temperature may be defined as
\begin{eqnarray}
  T_g=\frac{1}{2Nk_B}\displaystyle\sum\limits_{i=1}^N \left(m_i \left|
\bvec{v}_i - \bvec{\bar{v}}\right|^2 + I_i\left|\bsym{\omega}_i - \bsym{\bar{\omega}}\right|^2\right). \label{e:gtmp}
\end{eqnarray}

This quantity does not include, but is analogous to, the thermodynamic temperature \citep{walton1986}.  One important difference between a granular gas and a molecular gas is the inelastic nature of the collisions of the former, which leads to clumping and effectively serves to distinguish the behaviour of granular material.

\section{\textbf{THE ASTEROID SURFACE ENVIRONMENT}}\label{s:Environment}

By now it should be clear that asteroid surfaces are formed by regolith of various shapes, sizes, materials and, therefore, material properties.  In a granular aggregate, these material properties are intrinsically related to the size and shape of the grains, their atomic and electronic structure, and the gravitational field to which they are subjected, to mention the most important factors.  These properties will also play a role in how asteroids' surfaces, and asteroids as a whole, react to external agents such as gravitational fields of other planetary bodies, solar radiation pressure (Yarkovsky and YORP effects, particle transport and levitation) or impacts.  In what follows, we will explore these aspects of the surfaces of asteroids. Note that SI units should be assumed in all expressions in this section.

\subsection{\textbf{Surface characterisation}}

\paragraph{Materials -} Different mineral compounds form the regolith that is present in asteroid surfaces; they provide their spectral, thermal and some mechanical characteristics.  The first two are obvious as they have to do with the absorption, transmission and reemission of energy (Mazziero 2009); the third comes from how regolith is formed as that would be a reflection of the hardness of the material, the crystalline structure and forces between the surfaces of grains in contact (adhesion, cohesion and friction).

At the moment what is known about the materials that make up asteroids comes from the meteorites that have crashed on Earth, from spectral observations and, more recently, from the samples brought from asteroid Itokawa by the Hayabusa mission.  Through the research carried out on the available samples, it has been found that asteroids are formed mainly by pyroxine, olivine, plagioclase and iron compounds.  These materials have crystalline structures, and their detailed study belongs to the field of solid state physics or condensed matter physics.

\paragraph{Observed regolith characteristics -}  \sect{Observations} in this chapter has already summarised the main observations and interpretations made about the slopes and grain size distributions on asteroid surfaces.  

\paragraph{Surface gravity -} Up to this point, in the description and characterisation of the surface regolith of asteroids, there are no big differences with what can be found on Earth.  However, it is here where the similarities end as one of the most important factors affecting the dynamics of the regolith on asteroids is the ambient gravity; \ie the sum of the local gravitational field and centrifugal forces due to the rotation of the asteroid.  The calculated surface gravity of asteroids such as Itokawa and 1999 KW4 can be found in Yoshikawa et al. (this volume), Scheeres et al. (this volume), \cite{scheeres2010} and \cite{hartzell2013b}.

These surface gravity calculations show some important features that are not common in our terrestrial experience: (1) gravity is $10^{3}$-$10^{6}$ times smaller than Earth's gravitational field, $g$; (2) the gravitational field is not always perpendicular to the terrain; and (3) relatively small displacements on the surface of a small body could mean big changes in the gravitational field.  Among the main implications: escape speeds are in the order of cm/s; micro-meteorite impacts could transfer enough energy to generate surface or even global changes; stepping or landing on one of these aggregates could generate an ejecta field that could damage the instruments of a spacecraft or generate a local avalanche.

\paragraph{Friction -} Intuitively, the idea of friction is that of a force that resists the relative motion of two bodies that are in contact.  This resistance may appear in various ways; the work carried out by \cite{bowden1939a}, \cite{bowden1939b} and \cite{bowden1943} and later summarised by \cite{rao2008introduction} suggested that asperities or projections on the surfaces of the bodies adhere to form junctions.  Therefore, work must be done to deform and break these junctions, and this is accompanied by wear or erosion of material in the interfacial region.  Additional work is associated with the deformation of the material in a larger region near the interface (plowing).

The first attempt to formulate a macroscopic friction coefficient is attributed to \cite{coulomb1776}, who equated  it to the tangent of the angle of repose, by defining it to be the ratio of shear and normal stresses on an inclined pile of sand.  { The seminal work of \cite{bagnold1954} and \cite{bagnold1966} found that the frictional force varied as the square of the shear rate for grain-inertial flow in the regime of rapid shear.  On the other hand, \cite{midi2004} made it clear that the rheological properties of granular flows (friction, viscosity) depend on the shear rate \ie on the dynamics, thus putting the work of Bagnold and collaborators into context.}  

\cite{mehta2007granular} recognises that the proper microscopic formulation of inter-grain friction remains an outstanding theoretical problem.  In a granular material, it is not only the grain-grain surface friction that will determine the resistance of a grain to movement, or the resistance of the aggregate to be sheared and deform,  but also the grains' shapes, geometrical interlocking, packing and size distribution; all these factors are usually pulled together in a single term: the angle of (internal) friction that appears in the Mohr-Coulomb (MC) or Drucker-Praguer (DP) yield criteria.  

\begin{figure}[hbp!]
\begin{center}
\includegraphics[scale=0.33]{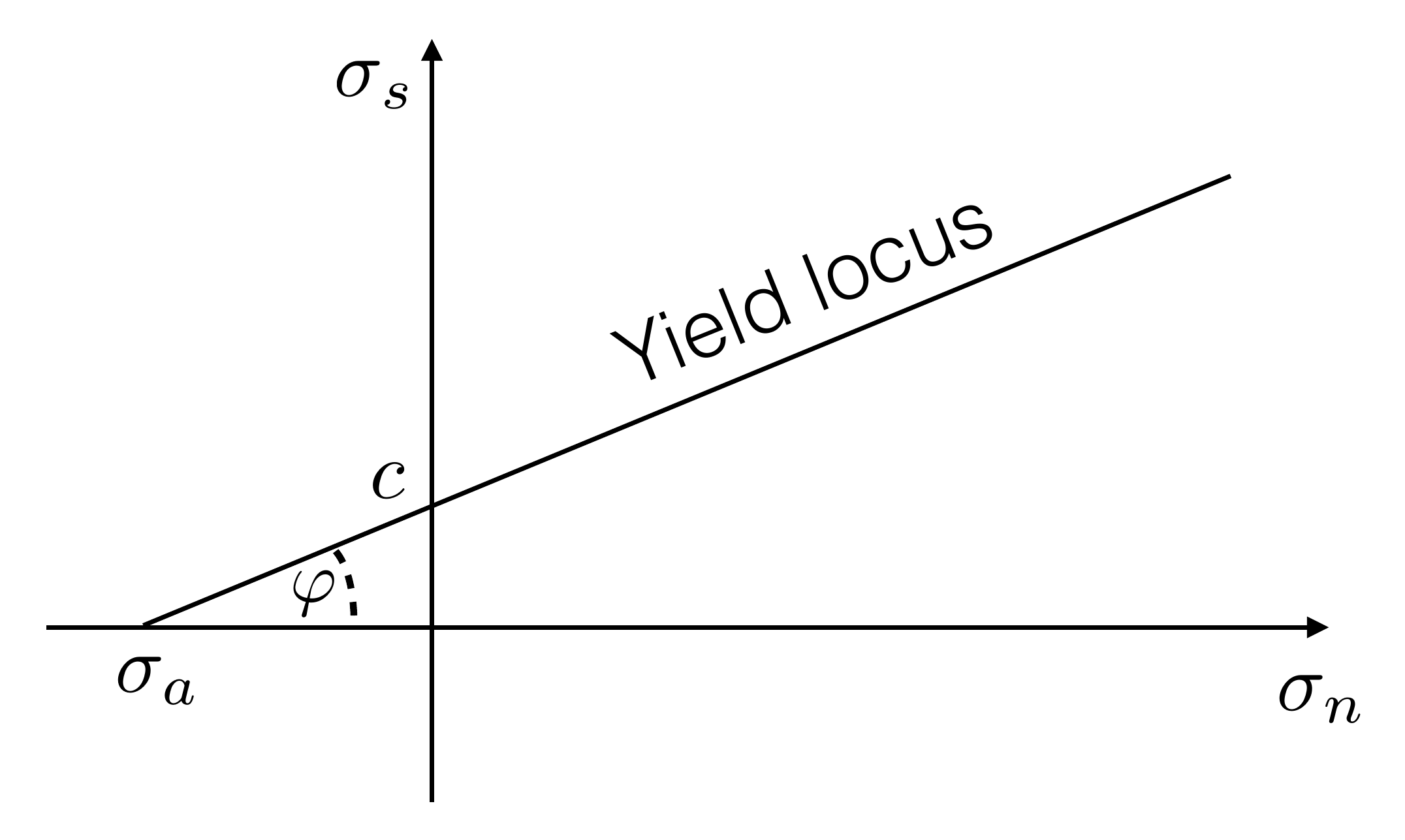}
\caption{Mohr-Coulomb yield criterion: normal stress ($\sigma_n$), shear stress ($\sigma_s$), cohesive strength ($c$), tensile strength ($\sigma_a$) and friction angle ($\varphi$).}
\label{fig:mohr-coulomb}
\end{center}
\end{figure}

{Within the Mohr-Coulomb yield criterion, the angle of internal friction is defined as the arctangent of the ratio of shear to normal compressive stress at the stability limit} (see \fig{mohr-coulomb}).  The MC criterion prescribes that shearing along any plane in a granular material cannot occur unless the shear stress ($\sigma_s$) on that plane reaches a value proportional to the normal (compressive) stress on that plane: $\sigma_s=\mu\sigma_n$. The proportionality constant $\mu$, the friction coefficient, is written as the tangent of the friction angle $\varphi$: $\mu=\tan(\varphi)$. Thus, the friction angle is a material property. The DP criterion is based on similar physical ideas, but uses a linear dependence of an average shear stress, as measured by the $J_2$ stress invariant, on the pressure \citep{holsapple2013, chen1988}.  Theoretically, if a material is cohesionless, the angle of repose corresponds to the angle of internal friction. On the other hand, for a cohesive aggregate, the angle of repose has to be such that is related to the cohesive strength of the material (see \cite{nedderman2005} for an in-depth explanation).

Cohesive forces (electrostatic and van der Waals are among the best known interactions) are kept in a different term and, within either criteria, do not affect the angle of friction, but only the angle of repose.  \fig{mohr-coulomb} shows the relation between the normal stress ($\sigma_n$), shear stress ($\sigma_s$), cohesive strength ($c$), tensile strength ($\sigma_a$) and friction angle ($\varphi$).  If cohesive strength is defined as the shear stress at zero normal stress then tensile strength is the normal stress at zero shear stress.

On Earth, a well-know experimental fact is that angles of internal friction of cohesionless granular materials vary from 25$^{\circ}$ for smooth spherical particles to 45$^{\circ}$ for rough angular particles \citep{carrigy1970,pohlman2006, kleinhans2011}.  Calculations made on asteroids Eros and Ida, and Martian satellites Phobos and Deimos show that  only Ida has more than 2\% (by area) of gravitational slopes above typical repose angles (35$^{\circ}$).  At this point it is worth explaining that the static angle of repose is the maximum slope that can be supported before the formation of an avalanche and a dynamic angle of repose is the slope that results after this avalanche has taken place.  The maximum angle of stability, critical angle and static angle of repose are the same; angle of repose and dynamic angle of repose are also the same. 
The static angle of repose may be related to cohesive forces including van der Waals forces, electrostatic forces and capillary forces in case of microscopic fluid pockets between the particles.  Using a parabolic flight experiment \cite{kleinhans2011} concluded that for decreasing gravity, the static angle of repose increases while the dynamic angle of repose decreased for all tested materials. 

\paragraph{Electrostatic forces and cohesion -} As mentioned before, any kind of cohesive force between the grains of an aggregate is going to increase the value of its angle of repose.  This includes van der Waals, capillary, electrostatic and magnetic forces.  The effects of these cohesive forces can be clearly seen in powders on Earth (flour, toner, pollen, chalk are common examples).  How they appear and when they are apparent in the behaviour of granular aggregates will be discussed in the next sections.

\subsection{\textbf{Cohesive and adhesive forces}}

\paragraph{Definition -} Cohesive and adhesive forces have the same origin, the only difference being that the term cohesive applies to the attractive force between molecules of the same material and the term adhesive applies to molecules of different materials.  For example, liquid water molecules attract one another and form water droplets with surface tension (this is cohesion); water molecules and silica molecules also attract one another and water can make a glass wet (this is adhesion).  These attractive forces are electromagnetic in nature and are appreciable when the electronic clouds of the atoms that form the surfaces of two bodies are within a few angstroms.  The term van der Waals forces is used here very loosely and refers to the totality of nonspecific attractive or repulsive intermolecular forces other than those responsible for ionic and covalent molecular bonds \citep{mcnaught1997}.  These interactions are often modelled by the Lennard-Jones potential \citep{jones1924}:
\begin{equation}
V_{LJ}=4\epsilon\left[\left(\frac{\gamma}{r}\right)^{12}-\left(\frac{\gamma}{r}\right)^6\right]
\end{equation}
where $\epsilon$ is the depth of the potential well, $\gamma$ is the finite distance at which the inter-particle potential is zero and $r$ is the distance between the two particles (neutral atoms or molecules).  Though other, more accurate forms of the potential exist, this one is usually used in computer simulations due to its simplicity.  The $r^{-12}$ term describes the Pauli exclusion principle due to overlapping electron orbitals and the $r^{-6}$ term describes the long-range attraction (van der Waals force, or dispersion force).  

\cite{johnson1971}, \cite{heim1999} and \cite{hughes2008} theoretically and experimentally studied the characteristics of the van der Waals force in granular materials. The cohesion between two spherical particles (radii $r_1$ and $r_2$) can be approximately described by \citep{castellanos2005, perko2001, rognon2008}:
\begin{equation} \label{e:Hamaker}
F_c = \frac{A}{48(t+d)^2} \frac{r_1 r_2}{r_1+r_2}
\end{equation}
where {\it A} is the Hamaker constant for the grains (4.3$\times 10^{-20}$ Joules for lunar soil), {\it t} is the minimum distance between the particle surfaces due to adsorbed molecules and $d$ is the width of any additional separation between the particles beyond that caused by the presence of the adsorbed molecules.  In the extreme environment of space the minimum distance between the materials can be much closer than possible on Earth where atmospheric gases, water vapor, and relatively low temperatures allow for significant contamination of surfaces ($d=0$).  These much cleaner surfaces and closer contacts allow for increased cohesion \citep{perko2001,scheeres2010}.

\cite{perko2001} define a cleanliness factor $S$ as $\Omega/t$, where $\Omega$ is the diameter of an oxygen ion ($O^{2-}$) and {\it t} is defined as above.  This being so, the cohesive force between a grain (radius, $r$) and a flat surface (or a much larger grain) is:
\begin{equation}
F_c=\frac{AS^2}{48\Omega^2}r
\end{equation}

\subsection{\textbf{Electrostatic forces}}

Electrostatic forces have been hypothesised to play an important role on the surfaces of asteroids, and have been specifically invoked as one means by which small dust grains can be transported across a body's surface. The first evidence of electrostatic lofting was the Lunar horizon glow observed by the surveyor spacecraft \citep{rennilson1974} at the terminator region.  A second discovery contributing to this hypothesis was the existence of ponds on Eros and other asteroids even though it has been found that their apparent distribution could have an observational bias.  Finally, there is also the fact that the Hayabusa mission was able to bring back samples of grains from asteroid Itokawa despite the malfunctioning of the sampling mechanism \citep{yano2006}.  It has been proposed that the electrostatic interaction between charged particles and a possibly charged sampler horn helped collect the sample \citep{tsuchiyama2011}.  Whether or not dust levitation occurs on asteroids is still an open question, although it is undoubtable that surface grains on these bodies are subject to electrostatic forces.  Unfortunately, as of yet, this is still not fully understood.

The electrostatic force felt by a particle on the surface of an asteroid is related to its location on the surface.  The charge density at any point on the surface is the result of the difference between the number of electrons that are deposited on it by the solar wind and those that leave the surface due to photoemission.  These two vary with the location of the surface and with time as the asteroid rotates and solar wind influences different areas of the surface.  Photoemission and solar wind interaction depend on the solar incidence angle and a variety of plasma-related phenomena that vary with solar longitude, respectively.  The resulting charge on the surface of the asteroid then influences the charging of the particle in question and influences the plasma environment (photoelectron and plasma sheaths) that will be experienced by the particle if it is lofted above the asteroid's surface \citep{scheeres2010}.

If grains are idealised as spherical, and we make the same assumptions as \cite{colwell2005} about the plasma sheet, it is then possible to demonstrate that for a particle of radius $r$ (surface area, $\Lambda = 4\pi r^2$), the electrostatic force that would provide lofting is:
\begin{equation}
F_{es}=\epsilon_0 E^2 \Lambda \approx 4\pi \epsilon_0 E^2 r^2\Rightarrow F_{es}\approx 9\times 10^{-9} r^2
\label{fes}
\end{equation}
where $\epsilon_0$ is the permittivity of vacuum and $E$ is the electric field.

Recent theoretical analysis and experiments have shown that cohesion will play a role in dust levitation \cite{hartzell2011,hartzell2013} and also that cohesion will dictate the electric field required for lofting for particles smaller than 1 mm on Itokawa (100 $\mu$m on Eros and 10 $\mu$m on the Moon).  Furthermore, these experiments have also shown that a balance between cohesive, gravitational and electrostatic forces is needed to ensure levitation.

\subsection{\textbf{The link between the surface environment and the geophysical features}}

In the previous sections we have tried to account for and describe the origin of the main forces that could affect grains on the surface of an airless planetary body.  However, an even more interesting aspect is the interplay of these forces as this determines the dynamics of the grains and the landscape of the surface as a whole.  With this in mind, and following the notation used by \cite{scheeres2010}, we define something called a {\it bond number} as:
\begin{equation}
B=\frac{F_c}	{W}
\end{equation}
where $F_c$ is the cohesive force acting on a grain and $W$ is its weight.  For an ambient gravitational acceleration of $g_A$ the ambient weight of a grain is defined as $W = mg_A$, where $m$ is the particle's mass.

On small planetary bodies, self-gravity between individual grains is much more important than on Earth and should be taken into account in calculations.  For two equal-size particles of radius $r$ and density $\rho_g=3500$ kg/m$^3$ (larger than the asteroid's bulk density) the bond number is:
\begin{equation}
B_{self}=G\frac{4\pi\rho_g}{3g_A}r \approx 1\times10^{-6}\frac{r}{g_A}.
\end{equation}

For the electrostatic force due to photoelectric emission alone, using eq.\ \ref{fes}:
\begin{equation}
B_{es}=6\times10^{-13}\frac{1}{g_Ar}.
\end{equation}
We note that triboelectric charging and other (not yet understood) mechanisms in the terminator regions could increase the electrostatic force to $F_{es}\approx0.1 r^2$.  This would, therefore, increase the electrostatic bond number to:
\begin{equation}
B_{es}\approx 7\times 10^{-6} \frac{1}{g_Ar}
\end{equation}

For cohesive forces and for material parameters of lunar regolith:
\begin{equation}
B_{c}=2.5\times10^{-6}\frac{S^2}{g_A r^2}
\end{equation}
where $S$ is the cleanliness factor defined earlier.

On small planetary bodies, for sufficiently small grain sizes, these bond numbers can easily attain values greater than 1, meaning that the grain's own weight can be overcome.  Strong cohesive forces give rise to highly porous structures first called ``{\it fairy castles}'' by \cite{hapke1963}.  Back then they attributed the existence of such structures to adhesive and long-range electrostatic forces that act between grains during deposition and influence their trajectories.  After that, the work of \cite{matson1983}, \cite{kreslavsky2003} and \cite{cassidy2005} shed more light on how these structures affect  photometric anomalies of the Moon.  Anomalous halos around small bright impact craters have been associated to changes in porosity probably related to some geologically recent damage of the equilibrium regolith structure.

{Additionally, the scaling of cohesive forces with ambient gravity means that cm-sized grains in a microgravity environment may behave as $\mu$m size grains in Earth's gravity.  Keeping this in mind, \cite{meriaux2008} and \cite{durda-epsc2013} have started research to understand the dynamics of cohesive powders under vacuum as a proxy to regolith-covered granular surfaces on asteroids.  Simulations carried out by  \cite{sanchez2014} and \cite{hirabayashi2014} have shown that modest values of cohesive strength (25 - 150 Pa) and a heterogeneous structure can drastically modify the maximum spin rates, disruption patterns and the existence or not of surface flow.  The results obtained by \cite{rozitis2014}, \cite{hirabayashi2014a} and \cite{scheeres2014} about 1950 DA and P/2013 R3 seem also to agree with the models, showing that values of cohesive strength under 100 Pa and angles of friction similar to those found on granular aggregates on Earth (35$^{\circ}$ - 45$^{\circ}$) are enough to explain the elevated spin rates of these asteroids.}

\section{\textbf{GEOPHYSICAL PROCESSES ACTING ON ASTEROIDS}}\label{s:Processes}

As we have already discussed, asteroids observed by spacecraft preserve records of geophysical processes that have operated at their surfaces and sometimes in their interiors. Here, these processes are classified into the following three categories: (1) exogenic phenomena, that are outer geophysical processes including impact cratering and slope failures/collapses, (2) endogenic phenomena, that are inner geophysical processes including ridges, faulting, and possible volatile and volcanic activities, and (3) other origins including tidal and YORP effects. Note that some processes dominantly acting on one asteroid might not necessarily act on other asteroids because the physical, especially mechanical, environments of asteroids vary significantly (for example, Vesta is considered to have been volcanically active because its mass is significantly large; 10 orders of magnitude larger than that of Itokawa, which is simply a pile of rubble). 

\subsection{Exogenic phenomena}

Because of the lack of an atmosphere, an asteroid is directly exposed to solar wind, cosmic and solar rays and influxes of meteoroids of varying sizes. An impact can largely modify the shape of an asteroid and even its arrangement (such as impact-induced break-up of an asteroid). This will not be discussed here, rather, we focus on surface processes resulting from impacts. 

On small bodies regolith is traditionally believed to result from repetitive impacts which excavate the surface and distribute ejecta materials, However, speeds of ejecta are typically greater than several tens of centimetres per second \citep{housen1979, housen2011}, which corresponds to the gravitational escape speed of kilometre-sized asteroids. { Impact debris reaccumulation, therefore, may not be solely responsible for the ubiquitous presence of regolith on small asteroids. Other regolith formation processes have been proposed including during contact-binary forming collisions of asteroids, by tidal forces, as well as the retention of regolith from a parent body \citep{barnouin2008, scheeres2007}. }  Using laboratory experiments and numerical simulations, \cite{delbo2014} have shown that thermal fragmentation induced by diurnal temperature variations breaks up rocks larger than a few centimetres more quickly than communition by micro-meteoroid impacts. The latter was demonstrated by adapting the lunar impact induced comminution rates of  \cite{hoerz1975} to asteroids. Because thermal fragmentation is independent of asteroid size, this process can also contribute to regolith production on larger asteroids. Production of fresh regolith originating in thermal fatigue fragmentation may, therefore, be an important process for the rejuvenation of the surfaces of NEAs \citep{delbo2014}. 

Once formed, resultant deposits of loose debris will be affected by the gravity in a longer timescale, where repeated disturbances cause overall slow motion in the downhill direction. The disturbances may be caused by processes such as small impacts and severe thermal cycling, while many of the surface processes are in some aspects similar to those terrestrial phenomena resulting from expansion and contraction processes, heating and cooling, wetting and drying, and freezing and thawing.  { We note that here there are many important temperature-related processes occurring on asteroid surfaces and these are discussed in Delbo et al. (this volume).}

\paragraph{Impact ejecta mantling -}

An impact will excavate the surface and create impact ejecta, which usually result in a deposit of debris surrounding the impact crater except for the cases in which the target has too little gravity to retain ejecta or too much porosity to produce it. The ejecta deposits normally affect at least the area within 5 crater radii by blanketing the original surface \citep{melosh1989}. In the case of gravity-dominated cratering, the thickness of the deposit, $H_b$, is given by 
\begin{equation}
H_b=0.14R_c^{0.74} [\frac{r_c}{R_c}]^{-3}, 
\end{equation}
where $r_c$ is the distance from the crater center and $R_c$ is the crater radius { \citep{mcgetchin1973}}. As suggested in this equation, the ejecta deposit is thickest at the crater rim and thins with increasing distance away from the crater. Impact-ejecta mantling may account for the absence of discernible surface features near craters on bodies such as Lutetia. 

When the ejecta deposit is continuous and clearly recognised to be the result of the cratering event, it is called an ejecta blanket. However, such a blanket is not recognised on the surface of a small asteroid. Considering the low gravity, as well as the often irregular shapes, the above equation might not be directly applicable for small asteroids. In fact, other than the local gravity, spin parameters, especially the rotational period, can significantly affect the situation and cause very asymmetric ejecta blankets as observed on Eros. For example, a three-dimensional SPH (Smoothed-Particle Hydrodynamics) simulation of a hemispheric-scale impact onto Vesta, which spins every 5.3 hours, show that variably shaped, multiply folded deposits can be formed \citep{jutzi2011} rather than a simple ejecta mantling. 

The presence of boulders adjacent to an impact site on Lutetia suggests that boulder generation is a common feature of large impacts on this asteroid \citep{sierks2011}. In fact, the fragments of ejecta deposits can be size-sorted through an impact event. Even though some interaction may occur between ejecta fragments in the denser parts of the ejecta curtain, general motion of the fragments is likely dominated by ballistics and thus follows a nearly parabolic trajectory above the asteroid before falling back to the surface. The size of the ejecta fragments near the base of the ejecta curtain is expected to be larger than the fragments higher in the curtain. 

One may wonder if ejecta may selectively escape from the surface of an asteroid never to return to mantle its original surface. This idea has been tested against the observations of Itokawa, for which 530 boulders larger than 5 m in size have been identified on its surface \citep{michikami2008}. Assuming the slope value on the cumulative plot (\fig{Cumulative}) from \cite{saito2006}, the cumulative number ($N$) of boulders may be approximated as:
\begin{equation}
N(>d) = 4.8 \times 10^4 d^{-2.8},
\end{equation}
where $d$ is the diameter of a boulder. If we assume that the above size distribution is continuous down to the size of a pebble, the volume of pebbles (4 mm to 6.4 cm in size) can be estimated as $1.9 \times 10^5$ m$^3$. Note, however, that smaller particles that exist on the asteroid surface are difficult to observe, because they are overlapped by larger boulders; the size distribution may, therefore, be different from that observed for larger boulders by 0.2-0.3 in the slope of log-log plot. This estimate is nonetheless of the same order as that estimated from the areas and depths of smooth terrains \citep[$2.3 \times10^5$ m$^3$; ][]{miyamoto2007}. It might, therefore, be appropriate to assume that no particular pebble-sized blocks have selectively escaped or accumulated \citep{miyamoto2014}.

\paragraph{Seismic shaking -}

Crater excavation resulting from an impact onto an asteroid is associated with a shock wave that severely shakes the terrain. Such shaking may induce local movements that cause a net downslope movement of loose surface material \citep{richardson2005} and this effect can be more important for a smaller asteroid because the seismic energy is unable to attenuate over a large volume.

Indeed, the first evidence for such seismic shaking on an asteroid was presented by \cite{thomas2005}. They showed that the formation of a relatively young crater (7.6 km in diameter) on asteroid Eros resulted in the removal of other craters as large as 0.5 km over nearly 40\% of the asteroid's surface. As burial by ejecta cannot explain the observed pattern of crater removal, and the areas with low small-crater densities correlate well with radial distance from the Shoemaker crater, they conclude that seismic shaking is the most probable mechanism. 

Assuming that the seismic energy is completely supplied by the kinetic energy of an impactor, the ratio of the maximum acceleration to the surface gravity, $a/g$, can be written as: 
 \begin{equation}
\frac{a}{g}  = \frac{3 f v_i}{G}\sqrt{\eta \frac{\rho_i}{\rho_a^3} \frac{D_i^3}{D_a^5}},
\label{e:aoverg}
\end{equation}
where $f$ is the seismic frequency in Hertz, $v_i$ is the velocity of the impactor, $G$ is the gravitational constant, $\eta$ is the seismic efficiency factor (the fraction of the original kinetic energy converted to seismic energy), $\rho_i$ is the bulk density of the impactor, $\rho_a$ is the bulk density of the asteroid, $D_i$ is the radius of the impactor, and $D_a$ is the diameter of the asteroid. For the case of a rubble pile, the seismic energy may attenuate significantly. In this case, a diffusive scattering theory as adopted by \cite{richardson2005} might be a realistic approach. In this theory, $a/g$ on a diffusive body may be written as in \eqn{aoverg} but multiplied by the scale factor $\exp (- \frac{f D_a^2}{K \pi Q})$, where $K$ is the seismic diffusivity and $Q$ is the seismic quality factor. Note that the exact values of some of the parameters for the above equations, such as $f$, $K$, $Q$, and $\eta$ are difficult to properly obtain. However, most importantly, both models show that the size of an asteroid is an important factor to determine the surface acceleration against the gravity; smaller asteroids vibrate easier than larger asteroids. Although it is difficult to constrain reasonable ranges of values for many parameters, the above equations generally suggest that {very small impactors may produce global shaking}. Thus, repetitive impacts on asteroids may cause significant shakings, believed to be responsible for the depletion of craters on Itokawa and small craters on Eros.

\paragraph{Mass movements -}

Once an asteroid is seismically shaken, reverberation continues until internal friction { and collisions} finally convert it entirely into heat. Similarly, some other disturbances such as heating and cooling may occur. When the formation of faults, rapid landslides, or other processes occur releasing horizontal stress, such disturbances generally make the ground surface expand perpendicularly to the slope. When the disturbance ceases, the ground surface contracts along the direction of gravity, which is not necessarily parallel to that of the above surface expansion (especially on a surface inclined against the local gravity). When such expansion and contraction occur cyclically, the materials covering the surface show overall migrations, which are essentially downslope displacements. This kind of downslope mass movements of dry, unconsolidated material is a common geological process on terrestrial planets { \citep{meunier2013}}. However, geophysically, those on terrestrial planets are generally interpreted in terms of the competition between gravity or inertia and inter-granular friction. Naturally for the case of a relatively smaller asteroid, the situation may be more complicated since other forces such as cohesive forces can play a significant role, as discussed in \sect{Environment}. 

A slow and cyclic creep process is not the only type of mass movement on an asteroid. More rapid examples include landslides, which take place when the acceleration due to { the ambient} gravity exceeds the ability of a rock on a slope to resist.  Roughly speaking, the conditions for this phenomenon to occur can be described by Coulomb's equation, $\sigma_s = c+\sigma_n \tan\varphi$ (see \fig{mohr-coulomb}). If the shear of a block of rock exceeds the maximum sustainable shear stress, the block slips and may be recognised as a landslide. This kind of mass movement can be found as small-scale, streak-like features such as observed on crater walls of Vesta (\sect{Observations}).  Indeed, mass movements can expose an interior, often recognised through differences in colour, as described in \sect{Observations}.

When a block of rock slips on a slope, sometimes the balance of shear stress and sustainable shear stresses inside the block changes. This causes a flow-like phenomenon, sometimes referred to as a debris flow or granular flow { \citep{legros2002}}. Some of the above-mentioned mass movements may be better explained by this process. 

When we consider a case in which a block of rock is resting on a slope at angle $\alpha$, $\sigma_s = (\frac{mg}{A}) \sin\alpha$, where $m$ is the mass of the block and $A$ is its basal area. Similarly, $\sigma_n = (\frac{mg}{A}) \cos\alpha$. Assuming the case in which the sliding rock is actually loose debris (a case in which cohesion is negligible in a terrestrial environment, and thus flow occurs rather than simply sliding), the largest shear stress in the block is achieved at the base of the rock. In this situation, $\sigma_s = \sigma_n\tan\varphi$, which gives the condition of $\alpha = \varphi$. In other words, in this case, the angle of repose is the same as the angle of internal friction and is independent of the gravitational acceleration.

The overall migrations of such flows are sometimes considered as gravity flows, whose speed $U$ may be described as $U \sim D(\rho g h)^{1/2}$, where $D$ is a coefficient of drag force, $\rho$ is the density of the flow, and $h$ is the thickness of the flow. The speed of the flow, therefore, depends on the square root of gravity { \citep{jop2006}}, indicating that the flow can be much slower than typical ones on Earth.

Another type of mass movement is electrostatic dust levitation (see also \sect{Environment}), which is proposed to be responsible for particle migrations on airless bodies exposed to both direct sunlight and the solar wind \citep{lee1996}. Smaller levitated particles on an asteroid may escape into space through the solar wind, but larger particles may settle back onto the surface. This may explain the smooth ponds on the surface of Eros \citep{robinson2001}, the formation of which clearly involves the settling of fines ($\ll$ cm-sized particles) in gravitational lows by a secondary process, after crater formation \citep[\sect{Observations}][]{cheng2002, robinson2002}.

\cite{cheng2002} suggest, however, that the pond material derives from the flanks of the bounding depression seismically shaken down to the bottom of the depression. Another idea for the formation of ponds is eroding of boulders; repeated day/night cycling causes material fatigue leading to erosion of the boulders \citep{dombard2010}. However, recent morphological analyses indicate that the deposited material most likely originates from a source external to the ponds themselves \citep{roberts2014b}. In addition, the morphology, geography, colour and albedo of the ponds { may be} consistent with formation by electrostatic levitation rather than seismic shaking  \citep{riner2008, richardson2005}. 

\paragraph{Regolith segregation -}

In addition to the size segregation that occurs during impact ejecta mantling, in granular flows where particles have different physical properties, particle segregation also occurs (as discussed in \sect{Grains}). One example of such segregation is in a granular avalanche. 
{ The dominant mechanism for segregation in granular avalanches is kinetic sieving \citep{Bridgwater1976} rather than effects of diffusive remixing, particle-density differences, or grain-inertia \citep{Thomas2000}. } As the grains avalanche downslope, there are fluctuations in the void space and the smaller particles are more likely to fall, under gravity, into gaps that open up beneath them because they are more likely to fit into the available space than the coarse grains. The fine particles, therefore, percolate towards the bottom of the flow, and force imbalances squeeze the large particles towards the surface \citep{gray2006}.   

A further example of regolith segregation, that may be responsible for the presence of large boulders on the surface of asteroids such as Itokawa and Eros \citep{asphaug2001, miyamoto2007}, is the ``Brazil-nut effect"  \citep{rosato1987}. The idea is that seismic shaking may cause the larger regolith particles to move up to the surface. { If this is the case, the interiors of rubble-pile asteroids having experienced this kind of evolution are likely to be composed of smaller particles than those observed at the surface. This may also lead to some variations of macro-porosity with depth inside the asteroid. }

The mechanism driving the Brazil-nut segregation is still under debate {\citep{Kudrolli2004}}. It has been suggested that the segregation { may} result from the percolation of small particles in a similar fashion to the kinetic sieving mechanism, but here the local rearrangements are caused only by the vibrations \citep[\eg][]{rosato1987, williams1976}. However, other experimental results from \cite{knight1993} have shown that vibration-induced size segregation { may} arise from convective processes within the granular material and not always from local rearrangements. 

Granular convection is, in fact, a process often invoked by the community of small-body scientists to interpret the surface geology of asteroids \citep{miyamoto2007,asphaug2007}. However, as discussed in \cite{murdoch2013b}, kinetic sieving and granular convection are strongly dependent on the gravitational acceleration \citep{thornton2005, murdoch2013a}. A weak gravitational acceleration { may, therefore,} reduce the efficiency of particle size segregation.  Indeed, recent numerical simulations and parabolic flight experiments of the Brazil-nut effect have shown that the speed at which a large intruder in a granular material rises is reduced as the external gravity decreases \citep{tancredi2012,guttler2013,matsumura2014}.  Therefore, all convective and particle segregation processes in a granular material on or near the surface of a small body may require much longer timescales than the same processes would require in the presence of a strong gravitational field

\subsection{Endogenic phenomena}

For terrestrial planets, ``endogenic processes'' are geological processes associated with energy originating from the interior of the planet, including tectonics, magmatism, metamorphism, and seismic activities. However, in the case of asteroids, the ultimate cause of tectonics or seismic activities may be difficult to clearly separate. 

Internal magmatism and volcanic processes are not often expected to occur on asteroids. However, differentiated early-forming asteroids should have experienced various kinds of volcanic activity, especially as a result of incorporating the heat-generating isotope $^{26}$Al. The howardite-eucrite-diogenite (HED) class of meteorites, whose parent body is believed to be the asteroid Vesta, have been studied in detail, and the magmatic origins of these rocks and their compositions as surface lavas or intrusions are well understood \citep[\eg][]{taylor1993}. Nevertheless, observations of Vesta show no conclusive evidence of volcanic features. The surface of Vesta, particularly the northern hemisphere, appears to be saturated with craters $>$10 km in diameter \citep{marchi2012}. This indicates that the surface regolith to depths of more than 1 km is significantly overturned. On the other hand, \cite{wilson2013} predicted typical lava flow dimensions are $\sim$10 m thick, much shallower than the thickness of overturned regolith, which might be the reason for the lack of clear evidence of lava flows or pyroclastics on the surface of Vesta. 

Even though volcanism is not a common process on an asteroid, some features indicate phase-changing of materials and their transportation to the surface. Possible melting flows identified on Vesta might be good examples. Also, more than 10 active (mass-shedding) asteroids have been reported \citep{jewitt2012} and the possible mechanisms for producing mass loss include dehydration stresses and thermal fracture. Although comets are not the focus of this chapter, we note that, in fact, strange features exist in the nuclei of both Tempel 1 and Wild 2. These features are interpreted as pits and scarp retreats resulting from venting of subsurface volatiles \citep{veverka2013}.

Some pitted terrains of Vesta (\fig{VestaSurface}(e)) are also believed to be related to subsurface volatiles; morphologic similarities of pitted terrains between Mars and Vesta suggest volatile release as an origin. However, because meteorites thought to originate from Vesta indicate low endogenic volatile content, it has been suggested that the volatiles have been delivered to Vesta's surface and are not endogenic. The source may be carbonaceous chondrites, which have been observed as clasts in howardites (meteorites most likely originating from the surface of Vesta), or perhaps comets. Later impacts into this water-bearing regolith would result in devolatilisation due to impact heating and melting \citep{denevi2012}.

Tectonic features are surface features created by internal stresses that fracture or deform the surface layer. Numerous tectonic features are documented on asteroids. For example, Vesta displays examples similar to those found on small Saturnian satellites such as Iapetus. Also, Rahe Dorsum on Eros is a ridge that extends for about 18 km around the asteroid, which resembles thrust fault structures on the terrestrial planets \citep{prockter2002}. Tectonic deformation cannot occur without forces to drive it. However, many sources of tectonic stress exist of both internal and external origin, and a clear separation and exclusive classification is difficult. 
 
For example, on Eros, the large ``spiral'' pattern southwest of Psyche, is suggestive of extensional tectonics whereas the spatial distributions of other ridge systems, as well as its estimated shear strength, indicate that the ridges are in fact thrust faults, which were formed by impact-induced compression \citep{watters2011,cheng2002, veverka2000}.  However, the ridge system on asteroid (2867) Steins (hereafter simply Steins), which is the most prominent feature recognised on its surface, may be formed through the { change in rotation rate due to the effect of solar radiation known as the YORP effect}. The effect spins Steins, making the surface seek for the object's potential-energy minimum, which may cause tectonic arrangements or landslide-like surface modifications towards the equator \citep{harris2009}. Either process can explain the formation of the ridge at the equator. A similar origin is proposed for an equatorial ridge of 1999 KW4 \citep{walsh2008}.

Grabens or linear depressions are found on many asteroids (see \sect{Observations}), which may seem strange given the fact that asteroids are covered by loose fragmental debris.  It is possible that these fractures are evidence of competent rock below the regolith. It has been suggested that they result from stresses from large impact events, which have refocussed and caused fracture far from the crater \citep{fujiwara1983, asphaug1996}, or that they are due to thermal stresses \citep{dombard2002} and/or body stresses induced by changes in spin. However, faulting can occur even in a granular matrix when it is cohesive relative to the applied stress. 
  
Grooves have been reported on Gaspra, Eros, Steins, Lutetia, and many other asteroids observed at high resolution. A subset of the grooves appear to be chains of pits (or crater-like indentations) in an almost linear arrangement. The global extent of a series of pitted chains found on Steins indicates these are not impact craters, because the chances of formation of many chains of these craters of similar size is highly improbable.  Instead, partial drainage of loose surface material into a fracture within stronger, deeper material is considered as a likely origin \citep{richardson2002, keller2010}. This explanation of how pitted grooves form was suggested to explain the features seen on Phobos \citep{thomas1979}. Experiments have demonstrated that, in such a model, the spacing of the pits along the groove is equal to the thickness of the regolith in which they form and is independent of regolith bulk density, grain size, shape, angularity and angle of repose \citep{melosh1989, prockter2002}.  Short and well-defined grooves can also be caused by a boulder that has bounced and rolled a short distance, but very few ($<$5) such tracks have been positively identified on the surface of Eros \citep{prockter2002, robinson2001}.

\subsection{Tidal and rotational effects}

There might be some morphological or even larger modifications of asteroids due to tidal forces. Modest influences are expected to include exposing layers of surface materials; to explain the fact that the laboratory spectra of ordinary chondrite meteorites are a good match to Q-type asteroids, \cite{binzel2010} and \cite{nesvorny2010} pointed out the possibility that Q-type NEAs underwent recent encounters with the terrestrial planets and the tidal force exposed fresh ordinary chondrite material on the surface. More considerable outcomes may include splitting the asteroid into a binary \citep{walsh2008a} or even catastrophically disrupting it in a manner similar to comet Shoemaker-Levy-9 at Jupiter. Such tidal effects by a terrestrial planet are considered as one of the most likely creation scenarios for asteroid families \citep{fu2005}. Also, strange shapes of some NEAs as revealed by radar images may have resulted from the tidal disruption processes and re-accumulations of disrupted fragments \citep[\eg][]{bottke1999}. 

YORP can modify both the rotation rate and the spin-axis orientation of small asteroids and has been identified as an important process driving their physical and dynamical evolution \citep{vokrouhlicky2003}. Asteroids of 5 km and smaller in radius in near-Earth orbits and a few tens of km in the inner main belt are subject to the YORP effect.  The spin-up of asteroids can have dramatic consequences \citep{scheeres2007,walsh2008}. For example, the deficiency in small craters on Steins is attributed to surface reshaping (through landslides) due to spin-up by the YORP effect \citep{keller2010}. The shape of the northern hemisphere of Steins is reminiscent of that of the NEA 1999 KW4, which has been attributed to spin-up by the YORP effect. A plausible scenario is that Steins was spun-up by YORP, leading to material sliding toward the equator to form the typical top-shape \citep{keller2010}. Mass shed from the equator of a critically spinning asteroid can accrete into a satellite \citep{walsh2008}. Alternatively, an asteroid may spin-up by the YORP effect until it reaches its fission spin limit and the components enter orbit about each other \citep{scheeres2007}. Asteroid pairs may be formed by the rotational fission of a parent asteroid into a proto-binary system, which subsequently disrupts under its own internal system dynamics \citep{pravec2010}. These binary asteroid formation mechanisms may explain the fact that asteroid pairs ubiquitously exist. For more information on asteroid binaries see Walsh et al. (this volume).

\section{INVESTIGATING REGOLITH DYNAMICS}\label{s:experiments}

Regolith processes on asteroid surfaces may not have ready terrestrial analogs. In order to study regolith dynamics in the unique asteroid environment described in \sect{Environment}, both experimental methods and numerical simulations can be used. 

\subsection{\textbf{Experimental methods}}

\paragraph{Creating reduced-gravity conditions - }

One of the major challenges for investigating the behaviour of regolith at the surface of an asteroid is to recreate the proper gravitational conditions. Microgravity, the condition of relative near weightlessness, can only be achieved on Earth by putting an object in a state of free-fall. Here we introduce some of the techniques used to perform experiments in reduced-gravity conditions.

Drop towers have been extensively used for microgravity experiments related to dust and regolith dynamics (\eg \citealt{hofmeister2009} studying granular flow under reduced-gravity, and \citealt{schrapler2012} and \citealt{beitz2011} investigating low-velocity collisions between dust agglomerates).  Parabolic flights can also provide a microgravity environment for regolith experiments  (\eg \citealt{murdoch2012a, murdoch2013a} performing experiments of granular shear and granular convection, \citealt{guttler2013} investigating granular convection and the Brazil-nut effect, and \citealt{dove2013} investigating particle charging and the dynamics of charged particles on the surfaces of airless bodies). Such flights are normally operated using modified commercial aeroplanes, however, smaller aircraft have also been used \citep[\eg][]{kleinhans2011}.  Other ways of achieving microgravity are using a sounding rocket (\eg \citealt{krause2004} studying the formation of dust agglomerates) or flying an experiment on the International Space Station (\eg \citealt{colwell2003} investigating low-speed impacts into dust). Further methods of simulating microgravity exist such as magnetic levitation and neutral buoyancy. However, to perform an experiment with particles using these techniques, every particle in the experiment would have to be neutrally buoyant or magnetically levitated. If, for example, only the experiment container is levitated or buoyant, the individual grains would still feel the gravitational field.

\paragraph{Creating the electrostatic environment - }

The electrostatic part of the asteroid environment is provided by the solar wind, which is essentially a stream of plasma (electrons and protons) that originates in the upper atmosphere of the Sun.  The energy of this plasma ranges between 1.5 and 10 keV.  Though there are many ways to obtain plasma, the one technique that is chiefly employed to study lunar and asteroid environments uses an emissive filament within a cylindrical stainless steel vacuum chamber.   Argon plasma is created by the impact ionisation using electrons emitted from a negatively biased and heated filament in the bottom of the chamber \citep{wang2012, hartzell2013}.  However, the experiments have a higher density and lower temperature than the solar wind.

Efforts are being made to improve these experimental techniques so that the obtained results are not only qualitatively, but also quantitatively correct, and so that they can be directly translated to the asteroid environment.  In spite of this, they carry size, time and cost constraints, and even in the best possible conditions, at times they fall short of the real environments that are the object of study.  These constraints are what make computer simulations, their development and understanding, attractive from a scientific point of view.  Computer simulations of course also have a cost as there are many simplifications and assumptions that have to be made (see \sect{num_methods}).  This means that there must be a trade off between their complexity and their realism.  A trade off that calls for a very careful look at the results and their interpretations as computational artefacts must be distinguished from fact.  If research is carefully conducted, simulations can be used to guide better experiments and predictions of simulations can be tested so that nature is understood.

\subsection{\textbf{Numerical methods}}
\label{s:num_methods}

With advances in computer hardware and software, numerical modelling
has become increasingly important to the study of granular systems in general,
and to granular systems in exotic environments such as the surfaces of small bodies, where conditions are difficult to replicate experimentally.
The different types of numerical
approaches can be divided into the broad categories of continuum
and discrete.
In the realm of numerical simulation, continuum approaches and
discrete approaches have their relative advantages and disadvantages that depend
on the specific investigation at hand.  In
general, discrete approaches attempt to treat material as individual
particles, sometimes with large particles as proxies for groupings of smaller
ones.  Continuum approaches average the physics of nearby particles,
and use smooth transitions to account for variance.
Continuum approaches are particularly well suited for high-speed collisions, where material phase changes and the finite propagation speed of sound waves are important.  In low-speed granular regimes, however, discrete approaches have the advantage of being able to capture the inherently discrete nature of granular systems and of being able to describe in great detail the properties of individual grains.  These properties are then used to solve for the frictional and cohesive forces that arise when grains come into contact with each other.  Discrete codes are also much better suited towards capturing the physics of slowly evolving granular systems such as those that take place on the surfaces of small bodies.

\subsubsection{\textbf{Numerical modelling of granular systems in planetary science}}
\label{s:lowgrav}

For many years, numerical continuum approaches have been used to address issues related to granular dynamics in the field of planetary science (\eg \citealt{holsapple1993}, investigating scaling laws for impact-induced catastrophic disruption, and \citealt{benz1994}, investigating different classes of two-body collisions).  Continuum approaches have since grown significantly in sophistication
(some in current use for the modelling of asteroid shapes and the scaling laws for disruption are,
\eg \citealt{holsapple2008,holsapple2009} and \citealt{sharma2009}).  Discrete numerical approaches, described below, have been in use in the field of planetary science since, \eg \citet{brahic1975,brahic1977}, who simulated Saturn's
rings, \citet{asphaug1994}, who simulated the breakup of comet Shoemaker-Levy-9 using a soft-sphere discrete element method (SSDEM), and \citet{richardson1998}, who conducted a more generalised numerical investigation into tidal breakups of small bodies using a hard-sphere discrete element method (HSDEM).  In light of advances in computer processor speeds, only quite recently have robust
versions of SSDEM begun to be applied to the realm of planetary science, and specifically
to the study of regolith dynamics in microgravity environments.  SSDEM granular physics codes are now developed or adapted specifically for planetary applications by various groups (\eg \citealt{wada2006}; \citealt{sanchez2011}; \citealt{schwartz2012}; \citealt{tancredi2012}) using various integration schemes and strategies to account for the types of friction between grains.  Other codes, using continuum approaches, have also been developed to investigate, for instance, collisions between porous aggregates \citep{sirono2004,jutzi2013}.  However, owing to the granular nature of the relevant dynamical processes involved, the regolith surfaces of small bodies are more commonly modelled using discrete element methodologies, specifically (although not exclusively) SSDEM.

\subsubsection{\textbf{ The continuum approach to regolith modelling}}
\label{s:continuum}

Continuum numerical modelling of granular material usually begins by defining
a systematic approach to averaging the physics across many particles (and
thereby treating the granular material as a {\it continuum}).  The approach typically
will involve dividing a parameter space or dimensional space into regions,
and then integrating the system forward in time.

{ In describing fluid mechanics equations for dense flows, relevant conservation laws are followed, often in a Navier-Stokes framework (\eg \citealt{haff1983}).}  At minimum, these conservation laws should include
mass conservation, momentum conservation, and the conservation of energy
together with the first law of thermodynamics.  These regions may be
described in Eulerian terms, where a volume in space is held constant,
with material passing in and out of this volume, or in
Lagrangian terms, where a region is described by the material itself as it
moves around in space (\eg see \citealt{springel2002} for a fully conservative
derivation of a Lagrangian treatment in a SPH code).  The numerical viscosity problems that stem from continuum
codes (a known problem since
\citet{vonneumann1950}, a result of the homogenising of material properties)
are somewhat easier to mitigate in
Eulerian approaches \citep{springel2010}, whereas the principle advantage
to the Lagrangian approach is that the resolution of the system adjusts
automatically to the movement of the material (see, \eg the \citealt{benz1994}
handling of two-body collisions).  Sophisticated codes that use
hybrids of Eulerian and Lagrangian descriptions, together with complex
physical laws and computational parameters, have been developed (see
\citealt{monaghan1988} for an early perspective).  In addition, there
have been significant advances in continuum coding approaches that
mitigate some of the problems of numerical viscosity, including sophisticated
differencing schemes (see, \eg \citealt{marti2014}).

In the modelling of
granular media, the continuum approach {often} treats the material as a
deformable solid and models it with some chosen finite-element
(\eg \citealt{crosta2009}, who make use of the definition of bulk plastic and elastic modulii) or
mesh-free (Lagrangian) method suited for the particular situation at hand
(\eg \citealt{elaskar2000}).
{
Stability problems (\eg the stability of granular piles or cliffs) require an elasto-plastic framework that defines, at minimum, some type of yield criterion (consider the simple 1-D coulomb yield criterion: conditions are static until the tangential force exceeds the product of the coefficient of static friction and the normal force).

Depending on the system, a continuum approach
could incorporate viscosity in some useful form (\eg \citealt{lagree2011}) and treat
the material as a fluid and use computational fluid dynamics (useful in describing outflows from, \eg crater walls).}
However, since successful simulations of asteroid surface dynamics entail the capturing of the
discrete nature of individual particles, the effects of such
homogenisation must be examined thoroughly.  \citet{haff1983}, in his article
describing his efforts to treat granular media as a fluid analytically,
considers many of the potential hazards and payoffs of using fluid dynamics from his
analytical approach.  These same considerations that arise analytically (\ie the
sharp boundary conditions on grain surfaces, including the complex frictional
forces at play on these surfaces), also arise numerically.

\subsubsection{\textbf{ The discrete approach to regolith modelling}}
\label{s:HSSSDEM}

The discrete-element method (DEM) is a general term applied to
the class of discrete approaches to the numerical simulation of particle motion,
where particles usually represent actual grains (or collections of grains), unlike
the continuum approach that uses averages to homogenise the material.
However, as continuum approaches use homogenisation schemes to simplify the complex and
rapidly varying physical quantities within a material, in discrete approaches, the physics within individual
particles are averaged, and thus the particles are defined only by their \textit{effective} behaviour,
described by quantitative parameters.  Nevertheless, these parameters are typically
borrowed directly from continuum mechanics, either by explicitly defining quantities such as the Poisson's
ratio and the Young's, bulk, and shear moduli, or by using derived quantities including spring constants and
friction coefficients.

{DEM collisional routines are typically built off of an $N$-Body routine.  In an $N$-Body framework, at the beginning of each timestep, forces
on each of the $N$ number of particles (bodies) in a given simulation are solved for and used to advance
the simulation ahead through time in small quantized \textit{steps}.  These forces can include, for example, external gravity or electromagnetic fields, and can incorporate the effects that the particles themselves have on the field (\eg interparticle gravity).  The collisional routines are then built on top of this framework, and define a new set of forces to account for the physical interactions that particles have with each other (of course these interactions are also electric at the molecular level).

In the standard implementation, particles are approximated
as having perfect spherical geometry (more complex geometries are also
possible).  Since DEMs tend to compute the motions of large numbers of individual
particles, it is relatively computationally intensive, which tends to
limit either
the length of a simulation or the number of particles in the simulation.}

\paragraph{The hard-sphere discrete-element method (HSDEM) - }
\label{s:HSDEM}

The numerical approach to solving the equations of motion in HSDEM is to
discretise the simulation in time, with variables progressing in small steps
(timesteps) by forward advancing along derivatives.  Collisions are predicted
in advance by analysing particle motion and
checking for potential contacts that may occur within the current timestep.
Particles are not allowed to penetrate each other (overlaps are not allowed).
HSDEM codes carry out collisions between spheres by treating collisions
as instantaneously occurring at a single point of contact that lies on the
particles' surfaces{; the sound speed through a particle is also instantaneous.}
Thus this methodology treats motions and mutual
interactions of non-deformable, indestructible (hard) particles.  The
assumption of hard particles allows collisions to be carried out
analytically, with post-collision velocities and rotations given by,
\eg \citet{richardson1994,richardson1995}.

\paragraph{The soft-sphere discrete element method (SSDEM) - }
\label{s:SSDEM}

SSDEM is commonly used in the study of granular
materials, and has
often been applied to industrial problems (\eg~\citealt{tsuji1992}; \citealt{cleary2002}; \citealt{kosinski2009}). 
The methodology has been applied in other disciplines of physics, such as chemical physics, under the
name of Molecular Dynamics (MD), where it is used to compute motions of atoms
and molecules and interactions between them (in fact, this application and nomenclature predates SSDEM's
use in granular physical contexts; \citealt{alder1959}).
In the complex case of simulating regolith dynamics, one must treat each of the
relevant frictional forces by generalising and applying the rules of interaction
between grains.
The basic methodology having been developed by \citet{cundall1979}, SSDEM treats macroscopic
particles as deformable spheres, allowing overlaps between
particles to act as proxies for actual deformation.  Particles are taken to be
in contact if and only if their surfaces are touching or mutually penetrating.  The
greater the extent of this penetration, the more repulsive force is generated.
The majority of codes either assume a linear force dependence or a Hertzian
dependence on penetration depth ($F\propto x$ or $F\propto x^{3/2}$,
where $x$ is the penetration in units of length).
Once a contact is established, particles are subject to frictional forces
often making use of material parameters based on continuum mechanical
theory;
these forces will vary depending on the specific SSDEM code
(see \citealt{radjai2011} for a comprehensive overview on the different
classes of SSDEM codes and common variations).

\subsubsection{\textbf{Benefits and drawbacks between these numerical approaches}}
\label{s:bendraw}

In contrast to HSDEM, where collisions are solved for analytically, based on the
positions and momentum states of the particles along with some
basic material parameters to describe the behaviour, SSDEM must resolve
each collision numerically.  As such, collisions typically require dozens
of timesteps to resolve.  In HSDEM, however, since
collisions are predicted in advance and then treated as instantaneous, it is
the external dynamics (\eg gravity) that drives the choice in step size rather
than the collision handling; although the timestep may
also be limited by concerns over missing a collision,
timesteps in SSDEM can often be smaller than
those used in identical HSDEM simulations by factors of 10$^{2}$.
In dense regimes, however, the speed of the
integration in HSDEM is typically limited by collisional bottlenecks owing to the fact that
collisions must typically be computed one-at-a-time in sequence, limiting
the efficiency of parallel processing.

{ During the finite amount of time that
it takes for two real particles to collide, the particles are in contact, exchanging energy
and momentum.  In sufficiently dense regimes, a third particle may intrude on this
collision by making contact with either particle or with both particles, changing the
outcome.  This exposes another drawback of HSDEM's treatment of collisions
between particles: multiple contact effects are not taken into account in HSDEM,
where collisions are separate and instantaneous.
Multi-contact systems of rigid, indestructible particles can, however, be solved using an algorithm
known as contact dynamics (CD), which treats these rigid particles as subject to Coulomb
static frictional forces \citep{moreau1994}.  HSDEM also must account for the problem of \textit{inelastic
collapse}, which occurs when a group of particles collides infinitely often in a finite time, causing
the simulation to grind to a halt (see, \eg \citealt{petit1987,bernu1990} for early numerical encounters of this effect,
and, \eg \citealt{mcnamara1992,mcnamara2000} for more complete quantitative descriptions).  Although sophisticated
collision-handling schemes have been tailored to help mitigate this problem \citep{petit1987,luding1998},
the simplest way to avoid the finite-time singularity
in HSDEM is straightforward: it requires setting some minimum impact speed or energy under which
the coefficient of restitution is unity (no dissipation).  This results in particles, even those in ``stable"
configurations, to always maintain some minimum energy state (temperature), which may not appropriately
capture certain low-energy granular regimes.  Problems of inelastic collapse do not arise in SSDEM or CD
methodologies.

SSDEM is nevertheless the appropriate choice over CD in very dense regimes of large numbers
of particles because CD must solve, through iteration between each timestep, the contact
forces between each particle in a contact chain.  Still, HSDEM and CD can be a more
appropriate collisional routine in more dilute regimes, where collisions
do not involve networks consisting of large numbers of particles (\eg higher energy, ``granular gas"
regimes; see \sect{Grains} of this chapter), and where sound propagation speed is unimportant
(the sound speed can be controlled in soft-sphere methodologies via a stiffness parameter).}
However, even in 2-body collisions,
HSDEM can make errors.  Particles can rotate significantly during realistic, finite, oblique
collisions, altering the outcome of the collision---an effect that does not occur between perfectly
rigid particles \citep{muller2012}.\hide{,
as is the fact that real collisions involve surface deformation
of the particles at the contact point along with complex contact forces
(friction) that are not well accounted for in HSDEM.}  In addition to CD, attempts have been
made to use HSDEM with added analytical corrections to account for rotations of the multi-body
system while particles are colliding \citep{muller2013}, and to account for finite collision
times (by ``pausing" collisions).  These are most effective
in regimes when (third-) particle intruders can be safely ignored.  Also, when two real (deformable)
grains just ``graze" each other, depending on the rigidity of the grains, they may interact
very weakly; however, these types of contacts
are given too much significance when using hard spheres (the assumption of grain
incompressibility in HSDEM and CD leads to the exchange of too much energy and momentum
during oblique impacts).

Despite { these} drawbacks, HSDEM and/or CD can be the appropriate choice in certain
dilute/ballistic regimes (\cf \citealt{richardson2011, murdoch2012}), where they are
advantageous over continuum models for their speed and accuracy, and often over
SSDEM for their speed given the ability to handle large timesteps.
These are regimes where collisional timescales may be long compared
to other dynamical timescales (the ``granular gas" regime; see \sect{Grains} of this chapter), where
contacts between grains do not persist and thus complex
frictional forces are less relevant, and where the propagation of disturbance waves (material sound
speeds) are unimportant.

For the simulation of dense environments, however, including many granular regimes
in which grain deformation, { finite sound speed,} multicontact physics, and the
complexity of { higher-order} frictional
forces during contact cannot be neglected, SSDEM is the better choice.
Although the use of small timesteps can limit its speed, it is well suited for true
parallelisation (without the
HSDEM drawback of having to compute collisions in serial order).
Presently it is possible to follow the evolution of millions of grains in close
contact and over a fairly large range of simulation conditions, something not
possible with HSDEM.

The search for contacts in SSDEM is a simpler task than
the search for contacts (collisions) in HSDEM.  Before
integrating over the next timestep, HSDEM must ask:\ will there be a
collision at any moment during the following timestep?  In contrast,
SSDEM needs only to ask if there are any overlaps presently occurring.  Effectively,
this means that contact searches are a 4-dimensional problem in HSDEM (3 spatial
dimensions and 1 temporal dimension) and a 3-dimensional problem in SSDEM (3 spatial
dimensions).  More complex wall boundary geometries are
more easily included in SSDEM (the SSDEM code implemented in \code{pkdgrav}, for example, allows
for a wider set of wall boundaries; these include the triangle, which can allow for
sophisticated 3-dimensional polyhedral shapes, along with those discussed in
\citealt{schwartz2012}).

As a direct comparison of the two DEM collisional methodologies, { HSDEM and SSDEM}, simulations of
low-speed rubble pile collisions were performed using both SSDEM
and HSDEM in the same numerical code \citep{richardson2012}.
In the tests, self-gravitating rubble piles (without friction or cohesive
forces) were collided together at low speed.
The results from the two collisional routines were generally
similar; SSDEM often, but not in all cases, showed a somewhat higher final
ellipticity of the largest collisional remnant, suggesting a higher shear strength
that may arise from its more careful treatment of contact forces and finite collisional
times.

\subsubsection{\textbf{The use of numerical simulation in the field of regolith dynamics}}
\label{s:DEM}

Several DEM numerical codes have been written with the specific aim of investigating and solving for regolith dynamics.
\citet{walsh2008,walsh2012} used the HSDEM collisional routine in \code{pkdgrav} \citep{stadel2001,richardson2000} to
study grain displacements and lofting due to YORP spinup.  Soft-sphere collisional methodologies have been used to
study regolith dynamics in low-gravity environments, which include subsonic impact cratering into regolith \citep{wada2006,schwartz2014}, the Brazil-nut effect \citep{tancredi2012,matsumura2014}, and regolith motion due to tidal forces \citep{yu2014}.
Also in the realm of asteroid surface science, several numerical investigations to study avalanche run-outs and angles of repose of regolith have been performed using both continuum codes (\eg \citealt{holsapple2013}, using a finite-differencing method) and using DEM (\eg \citealt{richardson2012dps}, using soft-sphere).

The inclusion of cohesion in numerical coding can be adapted to many different granular
dynamics applications in planetary science, including the study of regolith dynamics \citep[\eg in SSDEM:][]{schwartz2013,sanchez2014}.  Attractive interparticle forces may be used to treat ionic or covalent molecular bonds,
weaker intermolecular dipole-dipole bonds such as hydrogen bonds and London dispersion forces, or
electrostatic forces.

\section{\textbf{CONCLUSIONS}}

In this chapter we have presented a brief overview of the observations of granular surfaces of asteroids, our current understanding of the geophysical processes that may have occurred, and the state of the art experimental and computational methods used to study them and make new predictions.   The field of regolith dynamics in varying gravitational environments, including the study of near-Earth asteroids as self-gravitating aggregates, is a new field of planetary science that will continue to evolve with the development of better computational tools and experimental techniques, refinements in the theoretical models and new in-situ observations from up-coming space missions such as OSIRIS-REx (NASA) and Hayabusa 2 (JAXA).

\section{\textbf{ACKNOWLEDGEMENTS}}

We would like to thank D. C. Richardson and D. J. Scheeres for their comments on our chapter and L. Staron and O. Barnouin for their very helpful reviews.


\bibliographystyle{AstIV} 
\renewcommand{\bibname}{References} 
\bibliography{AsteroidsIV} 

\end{document}